\documentclass[notitlepage,prfluids,longbibliography,superscriptaddress]{revtex4-1}
\usepackage[
paperwidth=178mm, 
paperheight=254mm, 
bottom=15mm,
top=25mm,
left=20mm,
right=20mm]{geometry}
\usepackage{graphicx}
\usepackage{epstopdf, epsfig}
\usepackage[usenames,dvipsnames,svgnames,table]{xcolor}
\usepackage{libertine}
\usepackage{amssymb}
\usepackage{amsmath}
\usepackage{mathptmx} 
\usepackage[T1]{fontenc}
\usepackage[utf8]{inputenc}
\usepackage{newtxmath}
\usepackage{natbib}
\usepackage{microtype}
\usepackage{hyperref}
\hypersetup{
colorlinks=true,
linkcolor=blue,
urlcolor=blue,
filecolor=blue,
citecolor=blue
}

\makeatletter

\newcommand{\vect}[1]{\boldsymbol{#1}}

\newcommand\Fr{\mbox{\textit{F}}} 
\newcommand\im{\mathrm{i}\mkern1mu} 
\DeclareMathOperator\tr{tr}
\DeclareMathOperator{\sign}{sign}
\newcommand\Real{\mbox{Re}} 
\newcommand\Imag{\mbox{Im}} 

\begin{document}

\title{General linear stability properties of monoclinal shallow
    waves}

\author{Jake Langham}
\email[]{J.Langham@bristol.ac.uk}
\affiliation{School of Mathematics, Fry Building, University of Bristol, Bristol, BS8 1UG, UK}
\affiliation{School of Earth Sciences, Wills Memorial Building, University of Bristol, Bristol, BS8 1RJ, UK}
\author{Andrew J. Hogg}
\affiliation{School of Mathematics, Fry Building, University of Bristol, Bristol, BS8 1UG, UK}

\date{\today}

\begin{abstract}
We analyze the linear stability of monoclinal traveling waves on a constant
incline, which connect uniform flowing regions of differing depths. The classical
shallow-water equations are employed, subject to a general resistive drag term.
This approach incorporates many flow rheologies into a single setting and
enables us to investigate the features that set different systems apart.  We
derive simple formulae for the onset of linear instability, the corresponding
linear growth rates and related properties including the existence of monoclinal
waves, development of shocks and whether instability is initially triggered up-
or downstream of the wave front.
Also included within our framework is the presence of shear in the flow velocity
profile, which is often neglected in depth-averaged studies.  We find that it
can significantly modify the threshold for instability. 
Constant corrections to
the governing equations to account for sheared profiles via a `momentum shape
factor' act to stabilize traveling waves.
More general correction terms are found to have a nontrivial and potentially important
quantitative effect on the properties explored.
Finally, we have investigated the spatial properties of the dominant (fastest
    growing) linear modes. We derive equations for their amplitude and frequency
    and find that both features can become severely amplified near the front of
    the traveling wave. For flood waves that propagate into a dry downstream
    region, this amplification is unbounded in the limit of high disturbance
    frequency. We show that the rate of divergence is a function of the spatial
    dependence of the wave depth profile at the front, which may be determined
    straightforwardly from the drag law.
\end{abstract}

\maketitle

\section{Introduction}
Shallow flows of fluid, or other continuous media, are often modeled using a
pair of hydrostatic depth-averaged equations describing the conservation of
volume and the balance of streamwise momentum. Such models have been employed
in many different settings, including classical studies of turbulent open
channels~\cite{Jeffreys1925,Dressler1949,Craya1952,Whitham1974},
granular
flows~\cite{Savage1989,Forterre2003,Forterre2008,DiCristo2009,GrayEdwards2014},
mudflows~\cite{Ng1994,Liu1994} 
and gravity currents~\cite{Hatcher2000,Hogg2001}.
Specializing a shallow-layer model for each particular case often involves only the
selection of a constitutive law for material stresses, which does not affect the
mathematical structure of the governing equations.  In one spatial dimension,
these systems may be written generally in terms of the flow depth $h(x, t)$ and
depth-averaged velocity $u(x,t)$, as 
\begin{subequations}
\begin{gather}
    \frac{\partial h}{\partial t} + \frac{\partial~}{\partial x}(hu) = 0,\label{eq:mass dim}\\
    \frac{\partial~}{\partial t}(hu) + \frac{\partial~}{\partial
    x}\left[\beta(h,u)
    hu^2\right] + g_\perp h\frac{\partial h}{\partial x}  =
    g_\parallel h - \frac{\tau(h, u)}{\rho},\label{eq:mom dim}
\end{gather}
\end{subequations}
where 
$g_\parallel \equiv g\sin\phi$, 
$g_\perp \equiv g \cos\phi$, 
i.e.\
gravitational acceleration resolved parallel and perpendicular to the
local slope at angle $\phi$ to the horizontal (hereafter assumed constant),
$\beta$ is a corrective
shape factor that arises during depth-averaging (discussed below), $\tau$ 
models the basal drag on the flowing medium and
$\rho$ is the flow density, hereafter assumed constant.  By leaving $\tau$ as an
arbitrary function of the flow variables, many different shallow-layer
formulations may be analyzed collectively.  This approach was employed
previously by Trowbridge~\cite{Trowbridge1987}, who showed in the case of
$\beta(h,u) = 1$ that any spatially uniform shallow flow of depth $h_0$ and
velocity $u_0$ on a constant grade is
linearly unstable if
\begin{equation}
    \sqrt{g_\perp h_0} < \left|\frac{\tau(h_0, u_0) - h_0 \frac{\partial \tau}{\partial
    h_0}}{\frac{\partial \tau}{
    \partial u_0}}\right|
    = h_0 \left|
    \frac{\mathrm{d}u_0}{\mathrm{d}h_0}
    \right|.
\label{eq:trowbridge}%
\end{equation}
By using this inequality, stability criteria for particular systems may be
deduced with ease.  The right-most expression, which is particularly simple to
evaluate, is absent from the original analysis~\cite{Trowbridge1987} and
generalizes a stability criterion derived by Craya for turbulent water flows in
arbitrary open channels~\cite{Craya1952}.  We include its derivation as a
special case of our analysis in Sec.~\ref{sec:linear stability} and note
additionally that the inclusion of modulus signs in the inequality
permit the assumption of the positivity of its right-hand side to be
relaxed~\cite{Langham2021}. 

\begin{figure}
    \centering%
    \includegraphics{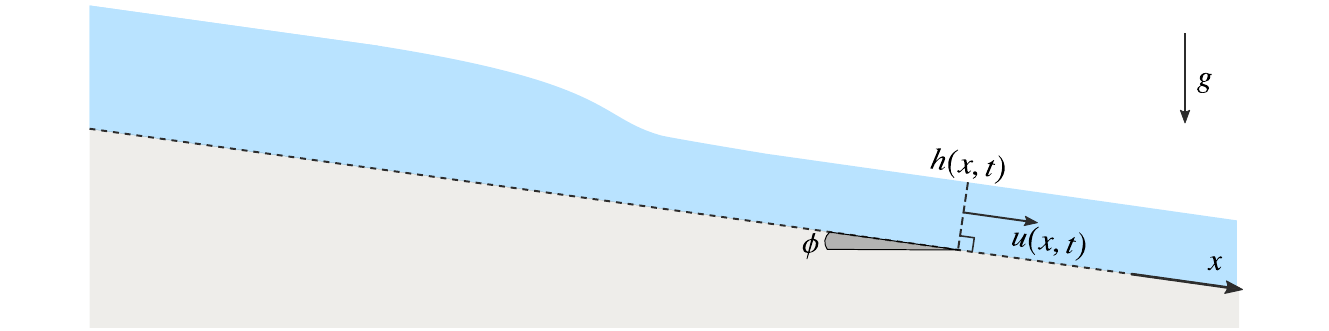}%
    \caption{%
	Diagram of the system under consideration, showing a flow of depth
    $h(x,t)$ and velocity $u(x,t)$, propagating down a fixed incline at angle
$\phi$ to the horizontal. The profile depicts a typical monoclinal traveling
wave solution connecting two uniform flowing layers.}
    \label{fig:setup}%
\end{figure}

The proliferation of shallow layer models in diverse settings makes the case for
conducting general analyses of this kind. Despite this, relatively few
studies have adopted a similar
viewpoint, see for example Refs.~\cite{Berlamont1981,Trowbridge1987,Zayko2019,Langham2021}.
The aim of this paper is to extend this program by considering the linear
stability properties of steady traveling wave solutions to Eqs.~\eqref{eq:mass
dim} and~\eqref{eq:mom dim}, with `monoclinal' depth profiles, which
monotonically connect regions with uniform flow depths far up- and downstream. 
This class of solutions
includes uniform layers as a trivial case and more broadly encompasses both
continuous and discontinuous fronts propagating between layers. 
A sketch of the system, indicating a typical monoclinal wave is given in
Fig.~\ref{fig:setup}.
Such states are experienced in nature as a surge between two shallow flowing
regions of different height.  They have been studied widely in the context of
turbulent open
water~\cite{Lighthill1955,Whitham1974,Jobson2001,Ferrick2005,Shome2006} and
more recently, in granular
flows~\cite{Pouliquen1999a,Gray2009,Razis2018,Razis2019,Kanellopoulos2021}.
The most mathematically extensive results on their stability are available in the
former case, where nonlinear stability theorems for monoclinal profiles have
been achieved~\cite{Yang2020,Sukhtayev2020}.
In the granular setting, the existence and stability of monoclinal
traveling were explored for a popular model that includes a
small diffusive regularization~\cite{Razis2018,Razis2019,Kanellopoulos2021}.
In addition to these cases, as shall be demonstrated below, monoclinal waves
are available as solutions to Eqs.~\eqref{eq:mass dim} and~\eqref{eq:mom dim}
for general drag formulations, provided that the closure permits the existence
of steady uniform layers. This includes established models where these states
have not been studied in detail.

Our study investigates the existence of traveling waves, their linear stability
and the spatial structure of the corresponding linear modes, within the general
setting of Eqs.~\eqref{eq:mass dim} and~\eqref{eq:mom dim}.  This consolidates
many existing results within a broader framework and provides a perspective
through which various properties of different systems may be understood.
Moreover, our conclusions may be simply applied in situations where individual
analyses have not been conducted.  In particular, for general drag laws we show
how to determine whether the monoclinal solution is continuous, discontinuous or even
admissible, as a function of downstream flow thickness relative to its upstream
thickness, the Froude number of upstream flow (defined shortly in
Sec.~\ref{sec:existence}) and potentially other parameters that determine the
resistance (Sec.~\ref{sec:existence}).  We compute the linear stability of both
continuous and discontinuous waves, determining properties of their associated
spectra and deducing a general criterion for instability (Sec.~\ref{sec:linear
stability}).  The fastest growing linear modes are shown to occur in the
asymptotic limit of high wavenumber (as in the case of uniform
layers~\cite{Trowbridge1987}) and general formulae for their growth rates are
given.  In certain cases, the frequency and amplitude of these modes are found
to be strongly amplified across the wave front.  Therefore, we investigate the
spatial structure of high-wavenumber modes and determine when, and why
amplification occurs using a WKB analysis~(Sec.~\ref{sec:eigenmode structure}).
It will be shown that this amplification is particularly extreme for wave fronts
that propagate into a region where there is no flowing material.  The resulting
analysis requires that the WKB approximation is asymptotically matched with
separate expansions for the behavior of modes near the wave front and ultimately
shows how the drag formulation dictates the rate of amplification with respect
to the wavenumber~(Sec.~\ref{sec:flood waves}).

Before proceeding, we note that the inclusion of the momentum shape factor
$\beta$ in Eq.~\eqref{eq:mom dim} also generalizes our analysis with respect to
many prior studies (including Trowbridge's analysis~\cite{Trowbridge1987}) of
shallow flow linear stability. It is defined as
\begin{equation}
    \beta(h,u) =
    \frac{1}{hu^2}\int_0^h \tilde{u}^2\,\mathrm{d}z
    =
    1+\frac{1}{h}\int_0^h\left(\frac{\tilde{u}}{u}-1\right)^{\!2}\,\mathrm{d}z,
    \label{eq:beta}%
\end{equation}
where $\tilde{u}\equiv\tilde{u}(x,z,t)$ denotes the velocity field prior to
averaging over the vertical coordinate $z$, i.e\ $hu = \int_0^h
\tilde u\,\mathrm{d}z$.  This parameter represents a
correction to the depth-averaging procedure used in deriving shallow-layer
formulations.  It is evident from Eq.~\eqref{eq:beta} that $\beta(h,u) \geq 1$. 
Most studies impose $\beta(h,u) = 1$, which formally corresponds to an inviscid
model of the flow with no shear in the velocity profile.
However, even small discrepancies from unity have been shown to have a
marked effect on solutions~\cite{Hogg2004}. We shall demonstrate that it also affects
their stability properties.  Since $\beta$ ultimately depends on the particular
flow rheology, as well as other observables such as the Reynolds number, we
leave it as a general function of $h$ and $u$ in our analysis.  While unknown
\emph{a priori}, we note that $\beta$ can be approximated for a given system
via Eq.~\eqref{eq:beta}, by employing an empirical steady-state representation
of $\tilde u$, such as the theory of Ref.~\cite{Reynolds1967}.

\section{Existence}
\label{sec:existence}%
We begin by postulating the existence of a traveling wave solution to
Eqs.~\eqref{eq:mass dim} and~\eqref{eq:mom dim}, propagating at wave speed $c_0$
and with constant depth $H$ and velocity $U$ in the far-field limit $x \to
-\infty$.  Hereafter, we refer to the limits $x\to-\infty$ and $x\to\infty$ as
the `upstream' and `downstream' directions respectively.  The various quantities
in the problem may be non-dimensionalised with respect to $H$, $U$,
$g_\parallel$ and $\rho$ using the transformations
\begin{subequations}
\begin{gather}
x \mapsto x g_\parallel / U^2,\quad
t \mapsto t g_\parallel / U,\quad
h \mapsto h / H,\quad
u \mapsto u / U,\quad \tag{\theequation a--d}\\
c_0 \mapsto c_0 / U,\quad
\mathrm{and}\quad
\tau \mapsto \tau / (\rho g_\parallel H). 
\tag{\theequation e,f}%
\end{gather}
\label{eq:variable subs all}%
\end{subequations}
%
A key control parameter in the forthcoming analysis will be $\Fr = U / (g_\perp
H)^{1/2}$. This dimensionless combination gives the Froude number of the flow
far upstream.

After non-dimensionalising, the governing equations~\eqref{eq:mass dim} and~\eqref{eq:mom dim} may
be rewritten in a more convenient frame by defining the coordinate $\xi = x - c_0 t$,
which follows the traveling wave.
On making this substitution 
and simplifying, a compact 
semilinear matrix equation may be obtained. 
We firstly give the resulting system for a general unsteady flow
$\vect{q}(\xi,t) \equiv [h(\xi,t), u(\xi,t)]^T$ in this frame. This is
%
%
\begin{gather}
    \frac{\partial \vect{q}}{\partial t} +
    J(\vect{q})\frac{\partial \vect{q}}{\partial \xi}
    = \vect{G}(\vect{q}),
    \label{eq:sw semilin}%
\end{gather}
where $\vect{G}(\vect{q}) = (0, 1-\tau/h)^T$ and $J(\vect{q})$ is
the Jacobian matrix, given by 
\begin{equation}
    J(\vect{q}) =
    \begin{pmatrix}
        u - c_0 & h \\
        \Fr^{-2} + B_2 & u - c_0 + B_1
    \end{pmatrix}.
    \label{eq:jacobian}%
\end{equation}
The terms $B_1$ and $B_2$ are placeholders for expressions which vanish when
$\beta(h,u) = 1$. They are
\begin{equation}
    B_1 = 2u(\beta - 1) + u^2 \frac{\partial \beta}{\partial
    u},\quad\mathrm{and}\quad
    B_2 = u^2 h^{-1}(\beta - 1) + u^2 \frac{\partial \beta}{\partial h}.
    \label{eq:B1B2}%
\end{equation}

The putative traveling wave is a time-independent solution of Eq.~\eqref{eq:sw
semilin}.  Substituting $\vect{q} \equiv \vect{q}_0(\xi) = [h_0(\xi),
u_0(\xi)]^T$
%
and integrating the first row of the resulting system gives
\begin{equation}
u_0 = c_0 + (1 - c_0) / h_0.
    \label{eq:u0}%
\end{equation}
Therefore, the steady velocity is a dependent variable, which may in turn be
substituted into the second row of Eq.~\eqref{eq:sw semilin},
via Eqs.~\eqref{eq:jacobian} and~\eqref{eq:B1B2}, to obtain:
\begin{equation}
    \frac{\mathrm{d}h_0}{\mathrm{d}\xi} = \frac{h_0 - \tau(h_0)}{h_0/\Fr^2 -
        (c_0-1)^2/h_0^2 + (c_0 - 1)B_1 / h_0 + h_0 B_2}.%
    \label{eq:h0 ode}%
\end{equation}
%
Since this is an ordinary differential equation in $h_0$ alone, the only bounded
traveling waves that may exist as solutions to Eq.~\eqref{eq:sw semilin} are either
everywhere monoclinal, or piecewise monotonic waves separated by discontinuities
(as in the case of a roll wave train, see e.g.\ Ref.~\cite{Dressler1949}).  For
continuous non-monotonic shallow waves to exist, tighter coupling between $h_0$
and $u_0$ is needed.  This is afforded by the presence of higher-order
derivatives (dispersion, diffusion) in some shallow layer formulations, see
e.g.\ Refs.~\cite{Johnson1972,Whitham1974,Razis2019}.

Our focus in this paper is purely monoclinal traveling waves.  Moreover, we have
assumed finite nonzero depth upstream, with $h_0(\xi), u_0(\xi) \to 1$ as
$\xi\to-\infty$, by our choice of non-dimensionalisation.
%
The downstream flow variables necessarily converge to constant values.
Therefore, we adopt the notation $h_0(\xi) \to h_\infty$ and $u_0(\xi) \to u_\infty$ as
$\xi \to \infty$ and note that the Froude number in this region is given by an
appropriate rescaling of the upstream value, $\Fr u_\infty h_\infty^{-1/2}$.
%
The far-downstream flow determines the speed of the traveling
wave, which we deduce from Eq.~\eqref{eq:u0} to be
\begin{equation}
    c_0 = \frac{1 - h_\infty u_\infty}{1 - h_\infty}.
    \label{eq:c0}%
\end{equation}
Since $u_0$ depends on $h_0$ and $c_0$ only, we note that $c_0 \equiv
c_0(h_\infty)$, with $c_0(0) = 1$ in the special case $h_\infty = 0$, 
where a wave front connects to a dry downstream region, referred to hereafter as
a `flood wave'.
We shall focus our analysis primarily on waves with $h_\infty
\leq 1$, since this is the most typically observed and studied case.

To illustrate our results and investigate the effect of drag, we
shall refer to various closures for the function $\tau$ throughout the text.
By considering Eq.~\eqref{eq:h0 ode} in the uniform flow regime far upstream, we see
that $\tau(1,1) = 1$. This
often allows at least one empirical parameter to be scaled out from a given
closure formula, leading to simple functional forms for $\tau$.
For example, turbulent fluid (Ch\'ezy) drag is given by $\tau(h,u) = u^2$,
while the drag on a viscously dominated fluid is $\tau(h,u) = u / h$.  A number
of results will be explored using the following rheology employed in the modeling
of granular media~\cite{Pouliquen2002,Jop2005,Jop2006}, 
\begin{equation}
\tau(h, u) = \frac{\mu(h, u) h} {\mu(1,
1)},\quad\mathrm{where}\quad
\mu(h,u) = \mu_1 + \frac{\mu_2-\mu_1}{1 + \zeta h^{3/2}/(\Fr u)},
\label{eq:pouliquen}%
\end{equation}
with $\mu_1$, $\mu_2$ and $\zeta$ empirically determined constants. Although
other possible parametrisations exist to describe granular flows via
specification of $\mu$ (examples include
Refs.~\cite{Savage1989,Pouliquen1999,Edwards2019}), our aim herein is not to
analyze the selection of individual closures.  Therefore, we simply fix the
illustrative values $\mu_1 = 0.1$, $\mu_2 = 0.4$, $\zeta = 10$ and refer to
Eq.~\eqref{eq:pouliquen} as `granular drag' throughout the paper. 

On specifying $\Fr$, $h_\infty$ and $\tau$, Eq.~\eqref{eq:h0 ode} may
be integrated to obtain a monoclinal wave solution.  In some cases, 
$\mathrm{d}h_0/d\xi$ is singular,
in which case a shock with velocity $c_0$ must be fitted at the singular point
to complete the wave profile.  Translational invariance permits us to locate
this at $\xi = 0$. Discontinuous solutions to Eq.~\eqref{eq:sw semilin}, must
satisfy the appropriate Rankine-Hugoniot conditions across a shock. These
ensure conservation of mass and momentum fluxes across $\xi = 0$ and are
straightforwardly obtained to be
\begin{equation}
    [h(u - c_0)]^{+}_- = 0 \quad\mathrm{and}\quad
    \left[hu(\beta u - c_0) + \frac{h^2}{2\Fr^2}\right]^+_- = 0,
    \label{eq:jump conds}%
\end{equation}
where $[f(\xi)]^+_- \equiv f(0^+) - f(0^-)$.
The downstream traveling wave is then given by $h_0(\xi) = h_\infty$, $u_0(\xi) =
u_\infty$ for $\xi > 0$, and at $\xi = 0^-$ we apply Eq.~\eqref{eq:jump conds}
to deduce that the height of the shock is
\begin{equation}
    %
    h_0(0^-) = 
    \frac{h_\infty}{2}\left\{
        \left[
            \frac{8\beta\Fr^2(c_0 - 1)^2}{h_\infty^3}
            +\left(
            1+
            \frac{2\Fr^2c_0^2(\beta-1)}{h_\infty}
            \right)^{\!\!2}
        \right]^{\!\frac{1}{2}}
        -1
    \right\}
    - (\beta - 1)\Fr^2c_0^2.
    \label{eq:h0minus}%
\end{equation}
Note that, since this equation does not bound the magnitude
of $h_0(0^-)$, it is possible for `monoclinal' shock solutions to be strictly
increasing for $\xi < 0$, before abruptly dropping to some $h_\infty < 1$ across
the shock. However, upturned shock waves of this sort are not necessarily stable.
We discuss the stability of such solutions in general later, in
Sec.~\ref{sec:stability regimes}.

\begin{figure}
    \centering%
    \includegraphics{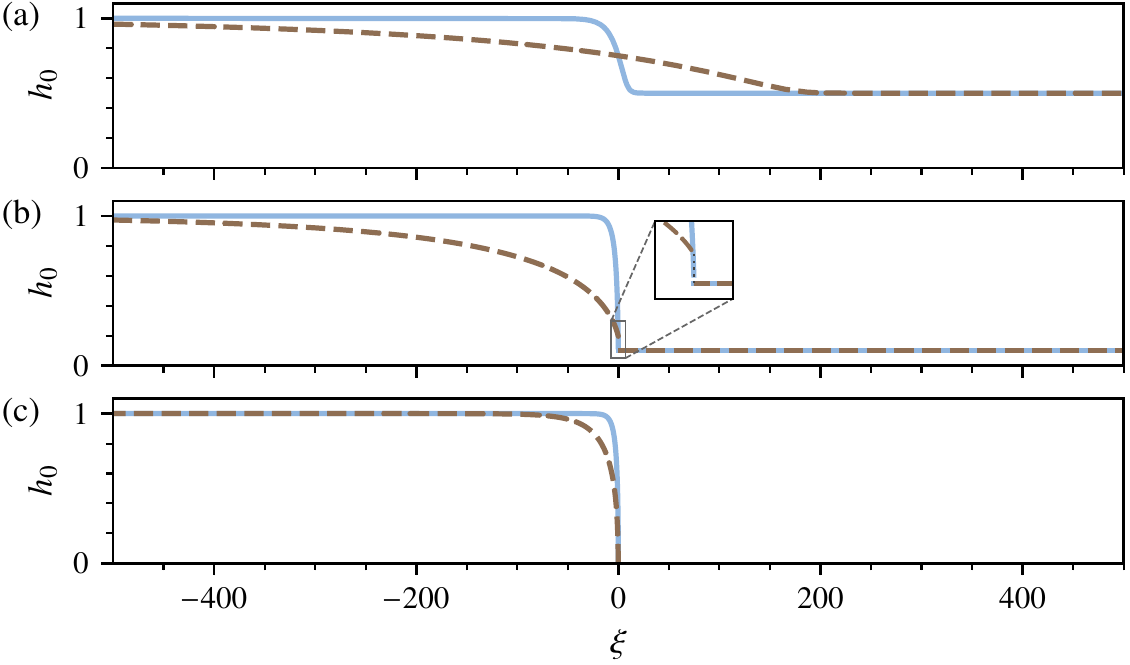}%
    \caption{Example traveling wave solutions satisfying Eq.~\eqref{eq:h0 ode}
        with $\beta = 1$ and $\Fr = 0.5$, for Ch\'ezy drag (solid blue) and
        granular drag (dashed brown). 
		The separate panels show connections from $h_0 = 1$
        upstream, to different downstream depths: (a)~continuous monoclinal
        waves with $h_\infty = 0.5$; (b)~shock profiles with $h_\infty = 0.1$;
        and (c)~flood wave solutions with $h_\infty = 0$.
    }%
    \label{fig:example tws}%
\end{figure}

Some example traveling waves are demonstrated in Fig.~\ref{fig:example tws}.
Across panels~(a)--(c), the solutions plotted
connect to progressively lower downstream levels. As $h_\infty$ decreases
from unity, continuous monoclinal waves develop a shock before
ultimately becoming flood waves when $h_\infty = 0$. 
To understand the regimes of Fig.~\ref{fig:example tws} in generality, it is
informative to consider the two characteristic curves $\lambda_1(h,u)$,
$\lambda_2(h,u)$ of the underlying system Eq.~\eqref{eq:sw semilin}, given by
the eigenvalues of the Jacobian defined in Eq.~\eqref{eq:jacobian}.  We compute
them to be
\begin{equation}
    \lambda_1(h, u) = 
    u - c_0 + \frac{B_1}{2}
-\sqrt{
\frac{h}{\Fr^2} + h B_2 +
        \left(
        \frac{B_1}{2}
        \right)^{\!\!2}
},\quad
    \lambda_2(h, u) = 
    u - c_0 + \frac{B_1}{2}
+\sqrt{\frac{h}{\Fr^2} + h B_2 +
        \left(
        \frac{B_1}{2}
        \right)^{\!\!2}
}.
    \label{eq:characteristics}%
\end{equation}
Note that if $B_2 < 0$, it is possible for the characteristics to be
complex-valued, leading to elliptic equations that cannot be well posed as
initial value problems. Therefore, we assume that $\lambda_1$ and $\lambda_2$
are distinct and real-valued, so that Eq.~\eqref{eq:sw semilin} is strictly
hyperbolic, as in the often used case with $\beta = 1$ (i.e.\ $B_1 = B_2 = 0$).  The consequences
of loss of strict hyperbolicity are addressed further in Sec.~\ref{sec:linear
stability}.
 
\begin{figure}
    \centering%
    \includegraphics{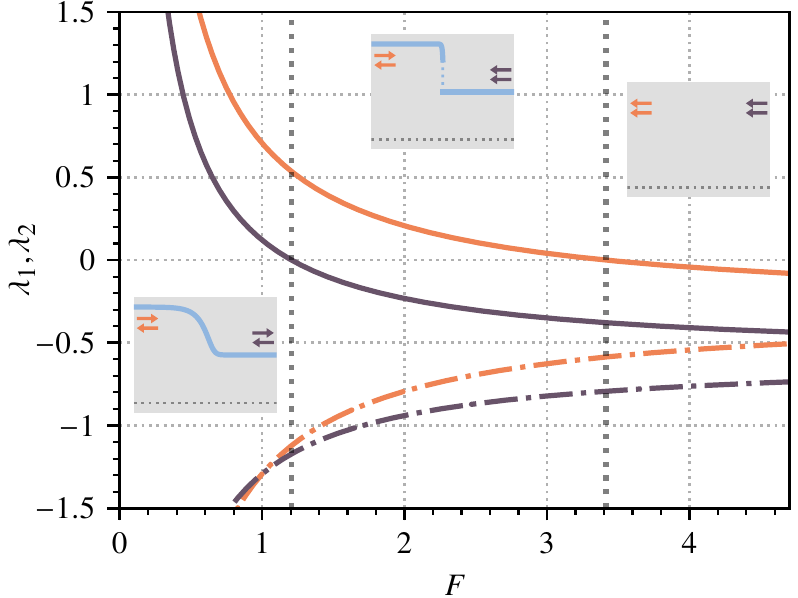}%
    \caption{Dependence of far-field characteristics $\lambda_1$ (solid) and $\lambda_2$
    (dash-dot) on $\Fr$, for Ch\'ezy drag monoclinal waves with $h_\infty = 0.5$
    and $\beta = 1$. Upstream values are plotted with orange curves, downstream
    with purple curves. The inlaid diagrams show the directions of propagation
    of the characteristics in the three distinct solution regimes given by
    inequalities~\eqref{eq:cont region} to~\eqref{eq:no monoclinal} (continuous
    waves, shock waves, no solution). These regimes are separated by vertical
    dotted lines. The wave profiles plotted in the leftmost
    and middle diagrams are $h_0(\xi)$ at $\Fr = 0.6$ and $1.8$ respectively,
    within the interval $\xi \in [-50,50]$.
    }
    \label{fig:characteristics}%
\end{figure}

The signs of $\lambda_1(h,u)$ and $\lambda_2(h,u)$ in the up- and downstream limits
$\xi \to \pm \infty$ dictate suitable boundary conditions for the problem and
ultimately, whether the two far-field regions must be connected via a shock.
We illustrate this with Fig.~\ref{fig:characteristics}, in
which the values of the upstream and downstream characteristics are plotted (in
orange and purple respectively) for Ch\'ezy drag and waves
with $h_\infty = 0.5$ and $\beta = 1$.
In the leftmost regime ($\Fr \lesssim 1.2$), the characteristics possess opposite
signs both up- and downstream and continuous monoclinal solutions exist
that connect the far-field regions. For greater values of $\Fr$, $\lambda_1$
changes sign in the far downstream. Consequently, both characteristics propagate
into the domain from the boundary at $\xi = +\infty$ and solutions contain a
shock (across which $\lambda_1$ changes sign) connecting the supercritical
downstream flow to the incoming wave.  When $\Fr \gtrsim 3.4$, $\lambda_1$
becomes negative in the $\xi\to-\infty$ limit also. The governing system can
no longer be posed with upstream boundary conditions in this regime, so
monoclinal solutions cease to exist.

%
%
%

We shall demonstrate that this picture does not qualitatively depend on the drag
law, or on the shape factor.  To determine the nature of the far-field
characteristics in general, we split our analysis into multiple cases,
since~$\beta$ is an unknown parameter, meaning that we cannot be sure of 
the sign of 
$u_0 - c_0 + B_1 / 2$ 
in either region.  Furthermore, for the remainder of this section only, we
make two simplifying assumptions.
First, we assume that $u_\infty(h_\infty)$ is a monotonically increasing
function.  This is true for most physical drag formulations, including all
examples given in this paper.  Therefore, $0 \leq u_\infty \leq 1$ when $0 \leq
h_\infty \leq 1$ and using Eq.~\eqref{eq:c0}, we conclude that $c_0 \geq 1$.
Second, we assume that the derivatives $\partial \beta / \partial h$ and
$\partial \beta / \partial u$ vanish, or are negligible, so that $B_1 =
2u(\beta-1)$ and $B_2 = (\beta -1)u^2/h$. Restricting these degrees of freedom
in this way permits us to find relatively succinct conditions for the different
solution regimes of the problem.

Firstly, we turn our attention to the upstream boundary, $\xi \to -\infty$, where
$u_0 - c_0 + B_1/2 = \beta - c_0$. 
Suppose that $\beta - c_0 < 0$. 
    Then $\lambda_1(1,1) < 0$.
    Moreover, 
$\lambda_2(1,1) < 0$, when
%
$\Fr^2 > [(c_0 - \beta)^2 - \beta(\beta - 1)]^{-1}$
%
is satisfied. In this case, both characteristics propagate out of the domain and
there are no admissible boundary conditions. For smaller values of $\Fr$, 
$\lambda_2$ becomes positive, so $h_0 \to 1$ may be imposed at
the upstream boundary.
Suppose instead that $\beta - c_0 > 0$. 
    This is the case for waves approaching the flood wave limit, $h_\infty \to
    0$ (where $c_0 \to 1$).
    Then $\lambda_2(1,1) > 0$.
    Furthermore, we can deduce that
    $\lambda_1(1,1) < 0$.
    To see this, suppose otherwise. Then, (recalling that $c_0 >1$)
    the following chain of inequalities would hold:
    \begin{equation}
        (\beta - 1)^2 > (\beta - c_0)^2 > \Fr^{-2} + \beta(\beta - 1) >
        \beta(\beta - 1),
    \end{equation}
    which contradicts $\beta > 1$.

Similar arguments apply in the case of the downstream region
$\xi\to\infty$, where we deduce that $\lambda_1(h_\infty, u_\infty)$ and
$\lambda_2(h_\infty, u_\infty)$ possess opposite
sign if and only if $\Fr^2 > h_\infty[(c_0 - \beta u_\infty)^2 - \beta(\beta -
1)u_\infty^2]^{-1}$ (otherwise both are negative).
%
%
Therefore, in summary, at both ends of the domain, the characteristics have opposite signs
if and only if
\begin{equation}
0 < \Fr < \left(\frac{h_\infty}{(c_0 - \beta u_\infty)^2 - \beta(\beta -
1)u_\infty^2}\right)^{1/2}.
    \label{eq:cont region}%
\end{equation}
(Depending on the sign of $\beta - c_0$, this inequality is satisfied
automatically in some situations.)
In this case, we anticipate a continuous monoclinal wave connecting from the
boundary condition $h = 1$ at $\xi = -\infty$, through to $h = h_\infty < 1$ at
$\xi = +\infty$.

If instead, we have
\begin{equation}
\left(\frac{h_\infty}{(c_0 - \beta u_\infty)^2 - \beta(\beta -
1)u_\infty^2}\right)^{1/2} < \Fr
< 
\left(\frac{1}{(c_0 - \beta)^2 - \beta(\beta -
1)}\right)^{1/2},
\end{equation}
then both characteristics at $\xi \to +\infty$ propagate into the domain.
The uniform layer in this region is therefore supercritical and connects to
the upstream flow via a discontinuity.

Finally, if
\begin{equation}
\Fr
> 
\left(\frac{1}{(c_0 - \beta)^2 - \beta(\beta -
1)}\right)^{1/2},
\label{eq:no monoclinal}
\end{equation}
then the characteristics propagate out of the domain at the upstream boundary
and there are no admissible boundary conditions there.  Note that, when $\beta =
1$, this inequality cannot be satisfied in the flood wave case, because $c_0 =
1$.  More generally, we can deduce from inequality~\eqref{eq:no monoclinal} that 
monoclinal traveling wave solutions exist for arbitrary $\Fr$, when the wave speed
$c_0$ satisfies
\begin{equation}
    1 \leq c_0 \leq \beta + \sqrt{\beta(\beta - 1)}.
\end{equation}
Since $c_0 \equiv c_0(h_\infty)$, this should be interpreted as an interval of
shallow downstream flow depths for which solutions always exist.

\begin{figure}
 \centering%
    \includegraphics{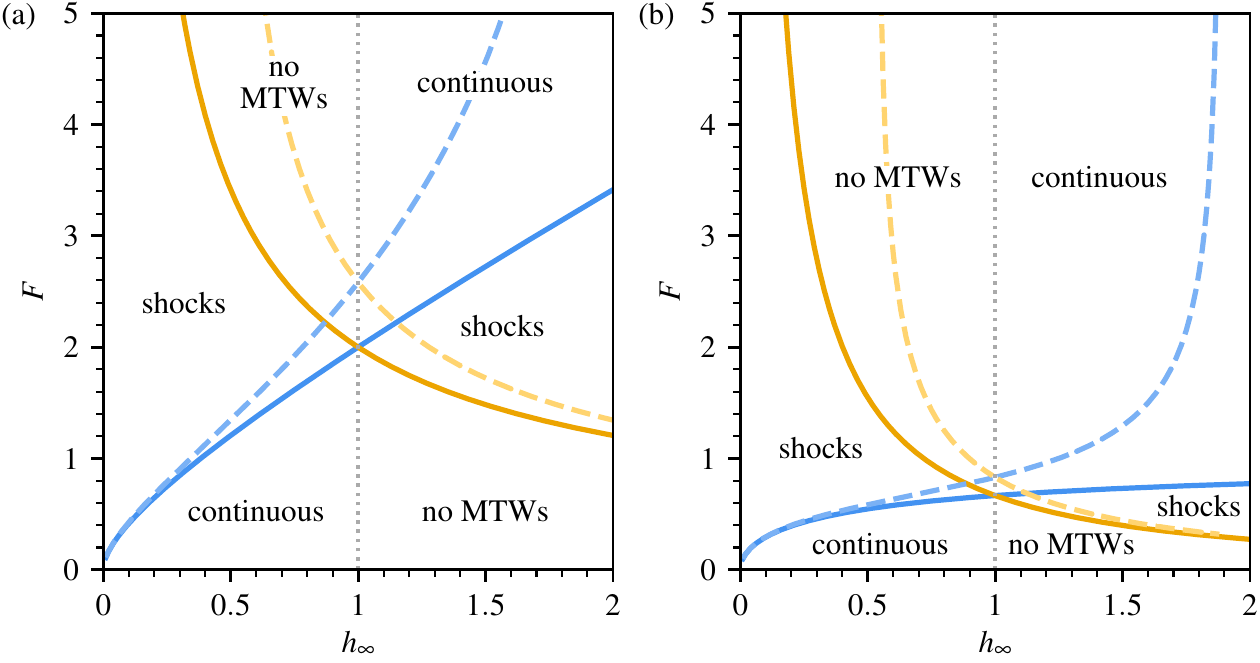}%
 \caption{%
     Existence of monoclinal traveling waves (MTWs) in the $(h_\infty,\Fr)$-plane for
     (a)~Ch\'ezy drag and (b)~granular drag.  The blue and orange lines denote
     the boundaries of the labeled regimes, given by the inequalities~\eqref{eq:cont region}
     and~\eqref{eq:no monoclinal} respectively.  Solid lines show the $\beta =
     1$ case and dashed lines show the $\beta = 1.05$ case in panel~(a) and
     $\beta = 1.2$ in panel~(b).
 } 
 \label{fig:existence}%
\end{figure}

In Fig.~\ref{fig:existence}, we indicate the solution 
regimes for both
(a)~Ch\'ezy drag and (b)~granular drag.
The standard case $\beta = 1$ is marked with solid curves, delineating the
boundaries of the inequalities~\eqref{eq:cont region} and~\eqref{eq:no
monoclinal}.  Increasing the vertical shear ($\beta > 1$) broadens the regions
of existence for both continuous and discontinuous solutions with $h_\infty <
1$.  For completeness, we continue the curves into the 
$h_\infty > 1$ parameter region.  In this case, the requirements for the
downstream layer to connect to the upstream are reversed, meaning that
continuous monoclinal waves exist at higher $\Fr$ than shock solutions.  For
uniform layers $h_\infty = 1$, there is no distinction between shocks and
continuous solutions, which exist for all~$\Fr$.  Consequently, the regime
boundaries merge. Note that the crossing point occurs at exactly the critical
$\Fr$ marking the onset of instability in a uniform layer (for example, $\Fr =
2$ for Ch\'ezy drag~\cite{Jeffreys1925} and $\Fr =2/3$ for granular
drag~\cite{Forterre2003}, when $\beta = 1$). In Sec.~\ref{sec:linear stability}
we will show why this must be true in general.

\section{Linear stability}
\label{sec:linear stability}%
To analyze the response of the traveling wave to perturbations, we introduce a
small amplitude disturbance that develops linearly with arbitrary complex growth
rate $\sigma$.  If the wave features a discontinuity, then we must allow for the
corresponding shock location to be perturbed also. Equivalently, we choose to
keep it pinned to $\xi=0$ and perturb the underlying coordinate frame.
Therefore, we write
\begin{subequations}
\begin{gather}
    \vect{q}(\xi, t) = \vect{q}_0(\xi) + \epsilon \vect{q}_1(\xi)\mathrm{e}^{\sigma t} +
    \ldots,\label{eq:q perturb}\\
    \xi = \xi_0 + \epsilon \xi_1 \mathrm{e}^{\sigma t} + \ldots,
\end{gather}
\end{subequations}
where $\xi_0 = x - c_0 t$ and the unknowns $\vect{q}_1$, $\xi_1$ are
$O(1)$ with respect to the small nonzero parameter $\epsilon$.
The corresponding perturbed shock velocity is $c_0 + \epsilon c_1
\mathrm{e}^{\sigma t}$, with $c_1 = -\sigma \xi_1$.
Linearising Eq.~\eqref{eq:sw semilin} with respect to this expansion 
and simplifying using the $O(1)$ expression,
we obtain the following compact equation governing the perturbation
\begin{equation}
    \sigma \hat{\vect{q}}_1
    + J(\vect{q}_0) \hat{\vect{q}}_1' 
    + N(\vect{q}_0) \hat{\vect{q}}_1
    = \vect{0},
    \label{eq:linear problem}
\end{equation}
where, using primes to denote total derivatives with respect to $\xi$, we have
defined a transformed perturbation vector
\begin{equation}
    \hat{\vect{q}}_1 = (\hat h_1, \hat u_1)^T \equiv 
    \vect{q}_1 + \xi_1 \vect{q}_0'
    \label{eq:q transform}
\end{equation}
and a matrix $N(\vect{q}_0)$ whose entries $N_{ij}$ are given by
\begin{equation}
    N_{ij} = \frac{\partial J(\vect{q}_0)_{ik}}{\partial q_j} (\vect{q}_0')_k
    - \frac{\partial \vect{G}(\vect{q}_0)_i}{\partial q_j}.
    \label{eq:N}%
\end{equation}
%
The variable transformation in Eq.~\eqref{eq:q transform} is typical of studies in
similar settings (e.g.\ Ref.~\cite{Yang2020}) and allows the linear equations to be
written in a form that is independent of whether shocks are present
($\xi_1 \neq 0$).  The only material difference between these two cases, is
that the evolution of any shock must also obey the jump conditions
of Eq.~\eqref{eq:jump conds} at $\xi = 0$, which after perturbing, become
\begin{subequations}
\begin{gather}
    \left[ h_1(u_0 - c_0) + h_0 (u_1 - c_1) \right]^+_- = 0,\label{eq:sh 1}\\
    \left[ h_0 u_0(\beta u_1 - c_1) 
    + (h_1 u_0 + h_0 u_1)(\beta u_0 - c_0) + 
\frac{h_0 h_1}{\Fr^2}\right]^+_- = 0,\label{eq:sh 2}
\end{gather}
\end{subequations}
to linear order.

In both far-field limits $\xi \to \pm \infty$, the base solution
$\vect{q}_0(\xi)$ is spatially constant.
Consequently, Eq.~\eqref{eq:linear problem} may be solved directly in these regimes
in terms of normal modes of complex wavenumber $k_-$ upstream and $k_+$
downstream.
Therefore, we seek eigenmodes of Eq.~\eqref{eq:linear problem} subject to the
boundary conditions
\begin{equation}
    \hat{\vect{q}}_1(\xi) \to \exp(\im k_\pm \xi)
    \hat{\vect{q}}_\pm
\quad
\mathrm{as}
\quad
\xi\to\pm\infty,
\label{eq:far bc}%
\end{equation}
where $\hat{\vect{q}}_-$ and $\hat{\vect{q}}_+$ are \emph{a priori} unknown
constant vectors.  
Then we may eliminate $\hat{\vect{q}}_1$ from Eq.~\eqref{eq:linear problem} in either of
the far-field limits, to obtain the dispersion relations
\begin{gather}
    \eta_\pm (\sigma + \im \lambda_{1\pm} k_\pm)(\sigma + \im \lambda_{2\pm} k_\pm)
    + \sigma + \im a_\pm k_\pm = 0,
    \label{eq:sigma quadratic}
\end{gather}
where (using a colon to denote the Frobenius inner product of matrices)
\begin{subequations}
\begin{equation}
\eta_\pm = \tr(N_\pm)^{-1}~\quad~\mathrm{and}~\quad~a_\pm = \eta_\pm\det(J_\pm)
    (J_\pm^{-1})^T
: N_\pm,
\tag{\theequation a,b}%
\end{equation}
\label{eq:nu a}%
\end{subequations}
with $\lambda_{1\pm} \equiv \lim_{\xi\to\pm\infty}\lambda_1$, $\lambda_{2\pm}
\equiv \lim_{\xi\to\pm\infty}\lambda_2$,
$J_\pm\equiv\lim_{\xi\to\pm\infty}J(\vect{q}_0)$ and
$N_\pm\equiv\lim_{\xi\to\pm\infty}N(\vect{q}_0)$.  In deriving
Eq.~\eqref{eq:sigma quadratic}, we made use of the fact that $\det(N_\pm) = 0$,
which follows from the definition of $\vect{G}$ and the fact that the first
term on the right-hand side of Eq.~\eqref{eq:N} vanishes as $|\xi|\to\infty$.
The relation holds for arbitrary $J$ in Eq.~\eqref{eq:sw semilin}.  It will be
useful for later discussion to appreciate that terms of the form $\sigma + \im
c k$ are the algebraic equivalent of the operator $\partial/\partial t +
c\partial / \partial \xi$ under the Laplace and Fourier transforms implicitly
employed in Eqs.~\eqref{eq:q perturb} and~\eqref{eq:far bc}. Hence,
Eq.~\eqref{eq:sigma quadratic} factorizes the linear dynamics of disturbances
in terms of wave transport operators with velocities $\lambda_{1\pm}$,
$\lambda_{2\pm}$ and $a_\pm$.



For any $\sigma$, there are two solutions of Eq.~\eqref{eq:sigma quadratic} for both
$k_-$ and $k_+$. The signs of $\Imag(k_-)$ and $\Imag(k_+)$ dictate whether the
corresponding eigenmode $\hat{\vect{q}}_1$ grows or decays in each far-field limit. Equivalently,
they determine the stability (with respect to spatial integration) of the fixed
point $\hat{\vect{q}}_1(\xi) = (0,0)^T$ of Eq.~\eqref{eq:linear problem} far up- and
downstream. Any linear mode must be bounded in order to be counted as
a small perturbation. For such a mode
to exist with a given $\sigma$, at least 
one solution of Eq.~\eqref{eq:sigma quadratic} for $k_-$ must have $\Imag(k_-) \leq 0$ and likewise, at least one
$k_+$ must satisfy $\Imag(k_+) \geq 0$. Note also that $\Real(k_-)$ and
$\Real(k_+)$ differ in general, indicating that the spatial modulation of
$\hat{\vect{q}}_1$ varies across the traveling wave front.  Examples of modes with
this interesting property are given below, in Fig.~\ref{fig:spec1}.

The constraints that Eq.~\eqref{eq:sigma quadratic} places on
the linear problem are illustrated in Fig.~\ref{fig:spec1}(a), which depicts the
spectrum for a monoclinal wave subject to granular drag, with $\Fr = 0.5$ and
$h_\infty = 0.5$.
\begin{figure}
 \centering%
    \includegraphics{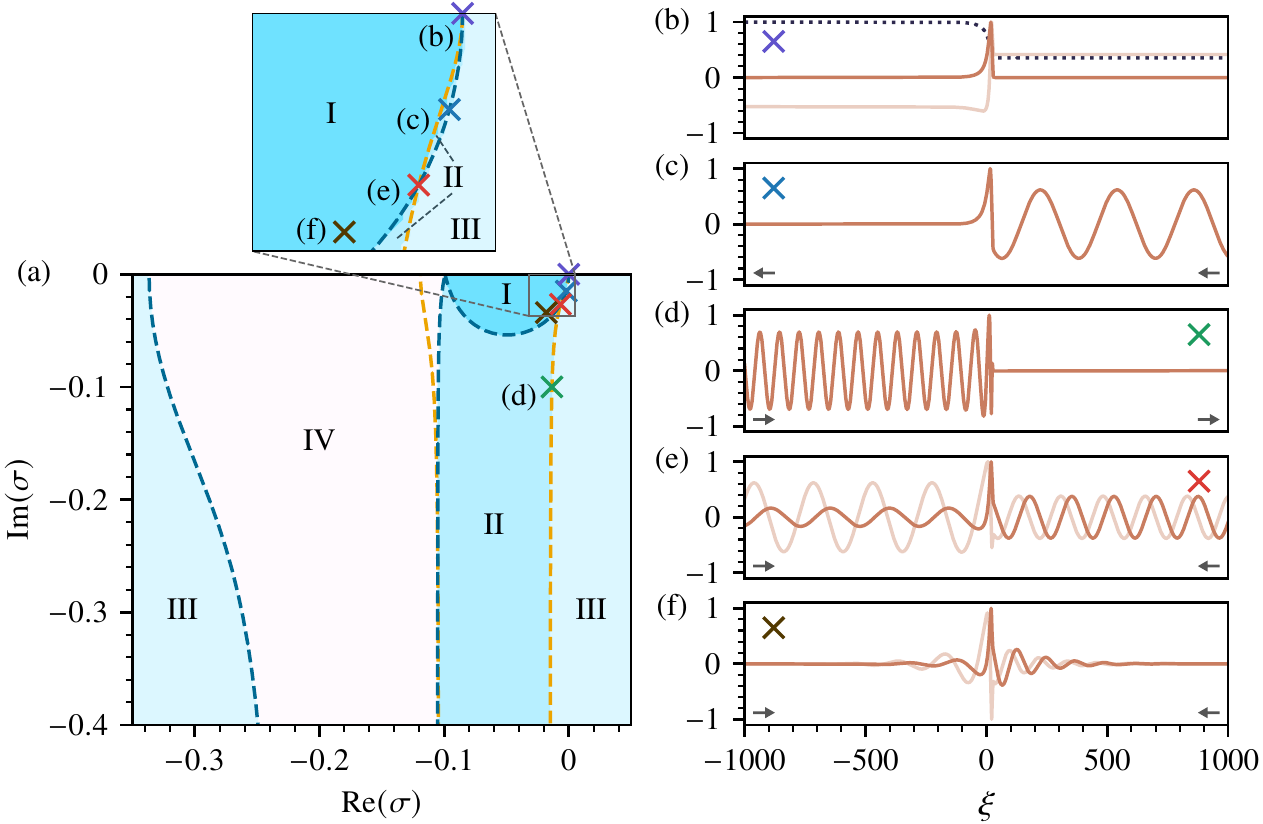}%
        \caption{%
        Spectrum diagram for the continuous monoclinal wave with $\Fr = 0.5$,
        $h_\infty = 0.5$ and the granular drag law given in
        Eq.~\eqref{eq:pouliquen}.  Different regions
        of the spectrum are indicated in~(a), labeled~I--IV according to the
        signs of the far-field spatial decay rates $\Imag(k_-)$ and
        $\Imag(k_+)$.  Details of this labeling are given in the main text.
        The critical lines $\Imag(k_-)=0$ and $\Imag(k_+)=0$ are plotted with
        yellow and blue dashes respectively.  Since the spectrum is symmetric
        about $\Imag(\sigma) = 0$, since $\exp(\im k\xi + \sigma t)$ is
        invariant with respect to the transformation $\sigma \mapsto
        \bar{\sigma}$,
        $k \mapsto -\bar{k}$ (where the overbar denotes complex conjugation), we have
        omitted the upper half-plane.  Regions~I \&~II comprise the essential
        spectrum, from which we show a selection of example modes,
        plotting~$u_1(\xi)$ with solid curves in panels~(b)--(f).
        Arrows indicate the direction of the perturbation velocities in the far
        field.
        In~(b), we
        also plot the base solution $u_0(\xi)$ (dotted).
        The exact growth rates of the computed modes, as indicated by crosses in
        panel~(a), are $\sigma =$~(b)~$0$,
        (c)~$-0.002-0.014\,95\im$, (d)~$-0.013\,5-0.1\im$,
        (e)~$-0.006\,678\,81-0.026\,7\im$ and~(f)~$-0.018-0.034\im$.
        Where two modes exist (b,e,f), the plotted curves are first
        orthogonalised with respect to the inner product $f \cdot g =
        \int_{-\infty}^\infty \overline{f(\xi)} g(\xi) \mathrm{e}^{-(\xi / 100)^2}
        \mathrm{d}\xi$.
    }%
    \label{fig:spec1}
\end{figure}
We find that the essential features of this figure are typical of other drag
rules (e.g.\ Ch\'ezy and viscous drag).
Regions of the plot are labeled according to the signs of
$\Imag(k_-)$ and $\Imag(k_+)$. 
In region~I, both $k_-$ solutions of Eq.~\eqref{eq:sigma quadratic} have
$\Imag(k_-) < 0$ and both $k_+$ solutions have $\Imag(k_+) > 0$.  That is,
$(0,0)^T$ is a repellor of Eq.~\eqref{eq:linear problem} at $\xi = -\infty$ and
an attractor at $\xi = \infty$. Crossing from region~I to region~II, there is a
sign change in one of the branches of $\Imag(k_+)$, indicated by the bounding
$\Imag(k_+) = 0$ curve (dashed blue).  This leaves a saddle point at $\xi =
\infty$.  For any $\sigma$ in either~I and~II, eigenmodes may be obtained by
integrating Eq.~\eqref{eq:linear problem} backwards in space from $\xi=\infty$
to $\xi=-\infty$.  Within region~II, bounded solutions to Eq.~\eqref{eq:linear
problem} must leave $\xi=\infty$ along the stable manifold of the far-field
fixed point, so the eigenspace is unidimensional in this case.  The union of
regions~I and~II forms the essential spectrum, whose modes are continuously
parametrised by $\sigma$.
Region~III designates the case where both up- and downstream limits are saddle
points.  In order for modes to exist in this region, the unstable manifold of
Eq.~\eqref{eq:linear problem} at $\xi\to-\infty$ must form a heteroclinic
connection with the stable manifold at $\xi\to\infty$.  Such connections are
not robust to perturbations of $\sigma$ and any modes in region~II are thus
isolated, forming the point spectrum of the linear operator.  Sophisticated
numerical methods exist for assessing the existence of these discrete
modes~\cite{Barker2018}.  However, they necessitate the specification of a
particular $\beta$ and $\tau$. Therefore, we must regrettably limit our scope to
considering the essential spectrum only. (See Sec.~\ref{sec:discussion} for
further discussion.)  In the current example of granular waves, we briefly
searched for modes in the unstable part of region~II various $\Fr$ and
$h_\infty$ and found none. 
Completing the qualitative description of Fig.~\ref{fig:spec1}(a), region~IV
indicates the regime where there can be no admissible eigenmodes, since either
$\xi=-\infty$ is an attractor, or $\xi=\infty$ is a repellor (or both).

We note that the essential spectrum of our example wave does not cross
$\Real(\sigma) = 0$ and is therefore linearly stable. In
Figs.~\ref{fig:spec1}(b)--(f) we plot a selection of modes from this region;
their locations in the spectrum are as indicated in Fig.~\ref{fig:spec1}(a).
(The velocity perturbation $u_1$ is plotted, but $h_1$ is similar in each case.)
Firstly, in panel~(b), which also includes the base solution $u_0(\xi)$ for
reference (dotted), we plot the two neutral stability modes with $\sigma = 0$.
The darker curve is (the velocity field of) the mode $(h_1,u_1) =
(-h_0',-u_0')$, which arises due to invariance of the traveling wave to shifts
along $\xi$. Its lighter counterpart corresponds to neutral perturbations along
the family of steady wave solutions that are parametrised by the downstream
depth $h_\infty$.  Both curves feature a sharp peak in the neighborhood of $\xi
= 0$ where the wave profile is steepest.
The mode in panel~(c) lies close to the origin on the
curve $\Imag(k_+) = 0$ and is consequently undamped in the downstream regime.
In the upslope direction, it decays rapidly towards a saddle point at $\xi =
-\infty$.  At the wave front, the mode features dramatic amplification, likely
inherited from the nearby neutral modes.
In panel~(d), we plot a mode on the curve $\Imag(k_-) = 0$, which is conversely
undamped in the upstream far field and decays rapidly as $\xi\to\infty$. For a
mode to be undamped in both directions it must lie on one of the discrete set of
intersection points of the critical curves. We isolate such a point in panel~(e)
and note that the corresponding modes are formed from the convergence of two
distinct wavenumbers.  Lastly, in panel~(f) we include a pair of modes in the
interior of the essential spectrum, which decay in both up- and downstream
directions. For each mode plot, we have included arrows showing the propagation
directions associated with the dominant perturbation velocities $-\Imag(\sigma)
/ \Real(k_\pm)$ in the far field regimes.  (From the two linear independent
components of each mode, we use either the least spatially damped wavenumber or,
in the case of saddle points, we take the component that remains bounded in the
relevant far-field limit.) We observe that the purely harmonic parts of
modes~(c)--(e) are directed towards the wave front in each case, regardless of
whether these undamped regions lie up- or downstream.

\begin{figure}
    \centering%
    \includegraphics{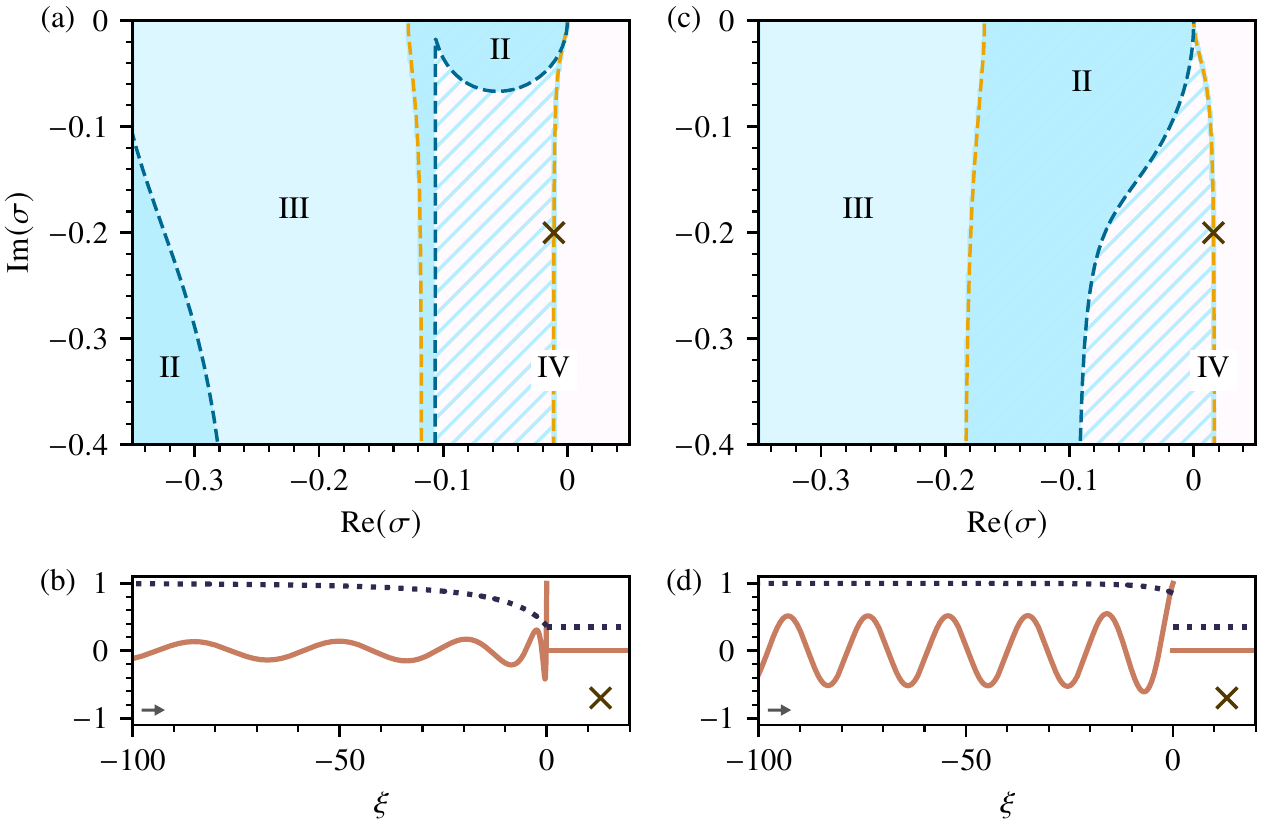}%
        \caption{%
            Spectra for discontinuous traveling waves with granular drag and
            $h_\infty = 0.5$.  Panel~(a) shows the case $\Fr = 0.55$, close to
            the boundary of, and within the regime of discontinuous states.  The
            labeling and color scheme matches the conventions of
            Fig.~\ref{fig:spec1}. An example mode is plotted in~(b), which shows
            the $u_1(\xi)$ field of the mode at the location labeled with a
            cross in~(a), $\sigma = -0.0109 - 0.2\im$ (solid).  The velocity of
            the base traveling solution, $u_0(\xi)$, is also shown (dotted).
            Panel~(c) plots the spectrum of an unstable case with $\Fr = 0.8$.
            In~(d), we plot the $u_1(\xi)$ field of the undamped unstable mode
            whose imaginary growth rate matches the example in~(b), in this case
            located at the cross in~(c), $\sigma = 0.0159 - 0.2\im$ (solid). The
            velocity of the base solution is also given (dotted).  In both
            panels~(b) and~(d), arrows indicate the direction of the
            perturbation velocity far upstream.
    }%
    \label{fig:spec2}%
\end{figure}

In Fig.~\ref{fig:spec2}, we demonstrate the changes to the spectrum
of the traveling wave as $\Fr$ is increased from its value of $0.5$ in
Fig.~\ref{fig:spec1}, with $h_\infty = 0.5$ remaining fixed.
Increasing $\Fr$ to $0.55$ [panel~(a)] leaves the wave slightly above the threshold for
shock formation [see Fig.~\ref{fig:existence}(b)]. The resulting spectrum is
qualitatively very different to the case of a continuous monoclinal wave.
This is because crossing the threshold between continuous and discontinuous
solutions induces a sign change in one of the branches of $\Imag(k_+)$, thereby
altering the stability of the fixed point at $\xi = \infty$. For example,
region~I in Fig.~\ref{fig:spec1} is labeled~II in Fig.~\ref{fig:spec2}(a),
since the attractor at $\xi = \infty$ becomes a saddle point when
solutions possess a shock. The sign change in $\Imag(k_+)$ occurs precisely because
of the sign change in $\lambda_{2+}$ that we argued in Sec.~\ref{sec:existence}
necessitates shock development. When $\lambda_{2+}$ passes through zero, the
leading term in Eq.~\eqref{eq:sigma quadratic} vanishes and one branch of
$\Imag(k_+)$ passes through a singularity associated with this degeneracy.  We
note that this also causes a topological change in the $\Imag(k_+)=0$ curve,
whose branch near the origin no longer forms a complete loop. Within the region
enclosed between the two branches of this curve (dashed blue), the
$\xi\to\infty$ point is a saddle, while to the left and right it is respectively
an attractor and a repellor. The stability of the $\xi\to-\infty$ point remains
qualitatively equivalent to the Fig.~\ref{fig:spec1} case.  Continuously
parametrised modes exist in region~II which is disconnected and stable.
Furthermore, in this case we may also consider modes which are identically zero
in either the upstream ($\xi < 0$) or downstream ($\xi > 0$) direction. In
particular, we find that upstream disturbances connecting to an unperturbed
downstream exist as bounded solutions to Eq.~\eqref{eq:linear problem}
everywhere enclosed within the two branches of $\Imag(k_-) = 0$ (dashed yellow).
The light blue hatched region denotes a region of the spectrum where only these
`upstream' modes exist. We include a plot of one of these modes in
Fig.~\ref{fig:spec2}(b) alongside the underlying traveling shock profile and
note that the mode amplitude is significantly amplified at the discontinuous
wave front.

Figure~\ref{fig:spec2}(c) shows the spectrum for an unstable traveling wave,
with $\Fr = 0.8$ and $h_\infty = 0.5$. In this case, we see that the regions
remain qualitatively similar to those in panel~(a). It is the upstream modes
only that cross $\Real(\sigma) = 0$, becoming unstable, with the maximum growth
rate attained for large values of $|\Imag(\sigma)|$, i.e.\ the most rapidly
propagating disturbances.  An example of such a mode is plotted in
Fig.~\ref{fig:spec2}(d), at $\Imag(\sigma)=-0.2$, matching the mode plotted in
panel~(b).  Modes with nonzero downstream perturbations are all stable at this
value of $\Fr$.  Following the continuation of our stability analysis below, we
shall show that this observation is not a generic property of these systems.
Whether the upstream regime is more vulnerable to instability than the
downstream or vice versa, depends on the particular drag closure.

\subsection{Deriving a general stability criterion}
\label{sec:deriving}%
The curves $\Imag(k_-) = 0$ and $\Imag(k_+) = 0$ where modes are undamped in the
far field necessarily dictate the boundary of the essential spectrum.
Therefore, to assess the stability of traveling waves (with respect to
continuously parametrised modes) in generality, we must determine when these
curves cross $\Real(\sigma) = 0$. Therefore, we suppose that $k_\pm\in\mathbb{R}$ for
the remainder of this section and write $\sigma = \sigma_r + \im \sigma_i$, with
$\sigma_r,\sigma_i \in \mathbb{R}$. On separating out the real and imaginary
parts of Eq.~\eqref{eq:sigma quadratic}, we obtain
\begin{subequations}
\begin{gather}
    \sigma_r^2 - \sigma_i^2 - \sigma_i(\lambda_{1\pm} + \lambda_{2\pm}) - k_\pm^2
    \lambda_{1\pm}\lambda_{2\pm} + \sigma_r \eta_\pm^{-1} = 0,\\
    \eta_\pm \sigma_r \left[
        2\sigma_i + (\lambda_{1\pm} + \lambda_{2\pm}) k_\pm 
        \right] + a_\pm k_\pm + \sigma_i = 0.
\end{gather}
    \label{eq:srsi eq}%
\end{subequations}
When $k_\pm = 0$, the solutions of these equations are $\sigma = 0,
-1/\eta_\pm$. The first of these corresponds to the aforementioned pair of
neutral stability modes, while the second corresponds to global perturbations
whose stability depends on $\sign(\eta_\pm)$.
We shall suppose that $\eta_\pm > 0$, which is the practical case of interest in
applications, where solutions are stable to spatially uniform modes.

Eliminating $\sigma_i$ and simplifying leads to the following expression
\begin{subequations}
\begin{equation}
    k_\pm^2 = \frac{\sigma_r(\eta_\pm \sigma_r + 1)(2\eta_\pm\sigma_r +
    1)^2}{\eta^3_\pm(s_1 - \sigma_r)(\sigma_r - s_2) (\lambda_{2\pm} -
    \lambda_{1\pm})^2},~~\mathrm{where}~~
    s_1\equiv \frac{a_\pm - \lambda_{2\pm}}{\eta_\pm(\lambda_{2\pm}-\lambda_{1\pm})},~
    s_2\equiv \frac{\lambda_{1\pm} - a_\pm}{\eta_\pm(\lambda_{2\pm}-\lambda_{1\pm})}.
    \tag{\theequation a--c}%
\end{equation}
\label{eq:ksqr}%
\end{subequations}
Therefore, when $|k_\pm| \gg 1$, we find that the two branches of $\sigma_r$
asymptotically approach the limiting values $s_1$ and $s_2$.  [For example, in
Fig.~\ref{fig:spec1}(a), these asymptotes are at $\sigma_r \approx -0.104,
-0.0149$ for $\xi\to-\infty$ and $\sigma_r \approx -0.231, -0.105$ for
$\xi\to\infty$, which match values computed for $s_1$ and $s_2$ in the
respective far-field regions.]
Moreover, for general $k_\pm$, Eq.~\eqref{eq:ksqr} may be differentiated to
obtain a formula for $\partial \sigma_r / \partial k_\pm$, which is zero if and
only if $k_\pm = 0$ or $\sigma_r = s_1, s_2$.
%
Hence, both branches of
$\sigma_r(k_\pm)$ are even functions which are monotonic with respect to $|k_\pm|$
and must be bounded by their values in the zero ($\sigma_r =
-\eta^{-1}_\pm, 0$) and high wavenumber ($\sigma_r = s_1, s_2$) regimes.

Since the growth rate is always stable for $k_\pm=0$, we conclude that
$s_1 \leq 0$ and $s_2 \leq 0$ for a linearly stable traveling wave, i.e.\ 
\begin{equation}
    \lambda_{1\pm} \leq a_\pm \leq \lambda_{2\pm},
    \label{eq:lam a}%
\end{equation}
in both far-field limits, with the onset of instability occurring at potentially
either of the critical cases $a_\pm = \lambda_{1\pm}$ or $a_\pm =
\lambda_{2\pm}$.  This recovers the results of
Whitham~\cite{Whitham1959,Whitham1974}, who noted that linear instabilities
occur in systems of conservation laws when the propagation velocity of the bulk
disturbance intersects with the characteristics. As noted earlier, the
quantities $a_-$ and $a_+$ may be identified as far-field wave velocities. They
correspond to a reduced description of the linear dynamics that omits the
propagation of high frequencies, which are carried by the leftmost term of
Eq.~\eqref{eq:sigma quadratic} at the characteristic velocities. Instabilities
first arise when the wave speeds of these low- ($a_\pm$) and high-frequency
($\lambda_{1\pm}$ and $\lambda_{2\pm}$) descriptions intersect (in either far
field).
 
Turning attention toward our particular application, we may
consult Eqs.~\eqref{eq:jacobian}, \eqref{eq:N} and~(\ref{eq:nu a}\hyperref[eq:nu a]{a,b}) and compute
\begin{subequations}
\begin{equation}
    \eta_\pm = \lim_{\xi\to\pm\infty} f_u^{-1},~~ a_\pm = \lim_{\xi\to\pm\infty} \left(
    u_0 - c_0 - h_0 f_h / f_u
    \right),
    \tag{\theequation a,b}%
    \label{eq:eta a specific}%
\end{equation}
\label{eq:eta a specific all}%
\end{subequations}
where we define the following terms, which arise from linearisation of
the drag function:
\begin{equation}
    f_h \equiv \frac{1}{h_0}\left( \frac{\partial \tau}{\partial
    h}\bigg|_{\vect{q}=\vect{q}_0} -
    \frac{\tau(\vect{q}_0)}{h_0}\right),
    \quad
    f_u \equiv \frac{1}{h_0}\frac{\partial \tau}{\partial u}
    \bigg|_{\vect{q}=\vect{q}_0}.
\end{equation}
By substituting Eqs.~\eqref{eq:eta a specific} into
Eqs.~(\ref{eq:ksqr}\hyperref[eq:ksqr]{b,c}) and making use of
Eq.~\eqref{eq:characteristics}, we find
formulae for the asymptotic (high wavenumber) growth rates
\begin{subequations}
\begin{equation}
    s_1 = \lim_{\xi\to\pm\infty} \left[ -\frac{f_u}{2} - \frac{1}{\lambda_{2\pm}
    - \lambda_{1\pm}} \left( h_0 f_h + \frac{B_1 f_u}{2} \right)\right],
    ~~
    s_2 = \lim_{\xi\to\pm\infty} \left[ -\frac{f_u}{2} + \frac{1}{\lambda_{2\pm}
    - \lambda_{1\pm}} \left( h_0 f_h + \frac{B_1 f_u}{2} \right) \right]\!.
    \tag{\theequation a,b}%
\label{eq:s1 s2}%
\end{equation}
\end{subequations}
These expressions can become unbounded if $\lambda_{2\pm} - \lambda_{1\pm} \to 0$.
This is a signature of ill posedness within the initial value problem
constructed in the linear stability analysis, which ultimately stems from loss
of hyperbolicity when the characteristics of the governing equations
coalesce~\cite{Joseph1990}.  While this is not possible within the classical
shallow-water framework where $\beta(h,u) = 1$, it could occur for other choices
of $\beta$. Specifically, from Eqs.~\eqref{eq:B1B2}
and~\eqref{eq:characteristics}, we see that $\lambda_1(h,u) = \lambda_2(h,u)$ if
\begin{equation}
    hu^2 \frac{\partial \beta}{\partial h} = -\frac{1}{\Fr^2} - u^2 (\beta - 1)
    - \left[
        u(\beta - 1) + 2u^2 \frac{\partial \beta}{\partial u}
    \right]^2.
\label{eq:ill p}%
\end{equation}
Increasing flow depth independently of other variables implies an increase in
flow Reynolds number, leading (in general) to blunter vertical flow profiles,
i.e.\ decreasing $\beta(h,u)$. Therefore, $\partial \beta/\partial h < 0$ is not
unexpected and particular care should be taken to avoid choices of
$\beta$ that lead to an ill-posed model.

Traveling waves can become unstable if $s_1$ or $s_2$ crosses zero in either
far-field limit. By rearranging either $s_1 = 0$ or $s_2 = 0$ we obtain the same
expression for the value of $\Fr$ at which flow becomes unstable in these
regions.  The lesser of these two values is the `critical' $\Fr$ above which,
linear instability of the traveling wave is guaranteed.
In the special case $\beta(h,u) = 1$, we label this
$\Fr_{\mathrm{Tr}}$ and find that
\begin{equation}
    \Fr_{\mathrm{Tr}}^2 = \min_{\ell = \pm\infty} \lim_{\xi\to \ell}
        \frac{f_u^2}{h_0f_h^2}.
\label{eq:FTr}%
\end{equation}
The square of the corresponding local Froude number is $\Fr_{\mathrm{Tr}}^2
u_0^2 / h_0$, where $u_0$ and $h_0$ in this instance are assumed to be evaluated
in the relevant far-field limit. Regardless of which limit applies, this agrees
(as it must) with the Froude number from Trowbridge's stability
criterion~\eqref{eq:trowbridge} for the linear stability threshold of a uniform
shallow layer.  In fact, since $u_0 \equiv u_0(h_0)$ and $\tau(\vect{q}_0) \to
h_0$ as $|\xi|\to\infty$, we may differentiate with respect to $h_0$ to deduce 
that
\begin{equation}
    \frac{f_h}{f_u} \to -\frac{\mathrm{d}u_0}{\mathrm{d}h_0}
    \quad\mathrm{as}~|\xi|\to\infty.
    \label{eq:fhfu lim}%
\end{equation}
Therefore, we obtain the slightly simpler and more intuitive formula 
\begin{equation}
    \Fr_{\mathrm{Tr}} = \min_{\ell=\pm\infty} \lim_{\xi\to\ell}
    \frac{1}{\sqrt{h_0}
    \left|
    \frac{\mathrm{d}u_0}{\mathrm{d}h_0}
    \right|
    },
    \label{eq:FTr 2}%
\end{equation}
which implies that the flow in the far fields is more unstable if the steady
velocity is more sensitive to changes in the flow depth.  As mentioned in the
introduction, this extends a much older result of Craya~\cite{Craya1952}, to
the case of arbitrary drag (and monoclinal waves).  While
$\mathrm{d}u_0/\mathrm{d}h_0$ is everywhere positive for most drag laws, the
modulus signs in Eq.~\eqref{eq:FTr 2} are required in general. For example,
confined channel geometries that narrow towards the top (such as partially
wetted pipes) can lead to a turning point in
$u_0(h_0)$~\cite{Camp1946,Chow1959}. Some additional implications of
Eq.~\eqref{eq:FTr 2} are given later in Sec.~\ref{sec:stability regimes}.

For general $\beta$, we shall write $\Fr = \Fr_c$ to denote the threshold of
instability. As before, we rearrange either $s_1 = 0$ or $s_2 = 0$ to obtain 
an expression for $\Fr_c$, which may be written in terms of
$\Fr_{\mathrm{Tr}}$, as so
\begin{equation}
    \frac{\Fr_c^2}{\Fr_{\mathrm{Tr}}^2} 
    = \min_{\ell=\pm\infty}\lim_{\xi\to \ell}\frac{1}{
    1 + \omega B_1
    h_0^{1/2}\Fr_{\mathrm{Tr}} - B_2 \Fr_{\mathrm{Tr}}^2
},
        \quad\mathrm{where}\quad
        \omega = \sign(f_u / f_h).
\label{eq:Frcrit}%
\end{equation}
Most typically, $\partial \tau(\vect{q}_0) / \partial h < 1$ and $\partial
\tau(\vect{q}_0) / \partial u > 0$, implying that $f_u / f_h < 0$ (as originally
assumed in Ref.~\cite{Trowbridge1987}).  Nevertheless, Eq.~\eqref{eq:Frcrit}
accounts for choices of $\tau$ where these inequalities do not necessarily hold.
%

\begin{figure}
 \centering%
    \includegraphics[width=\textwidth]{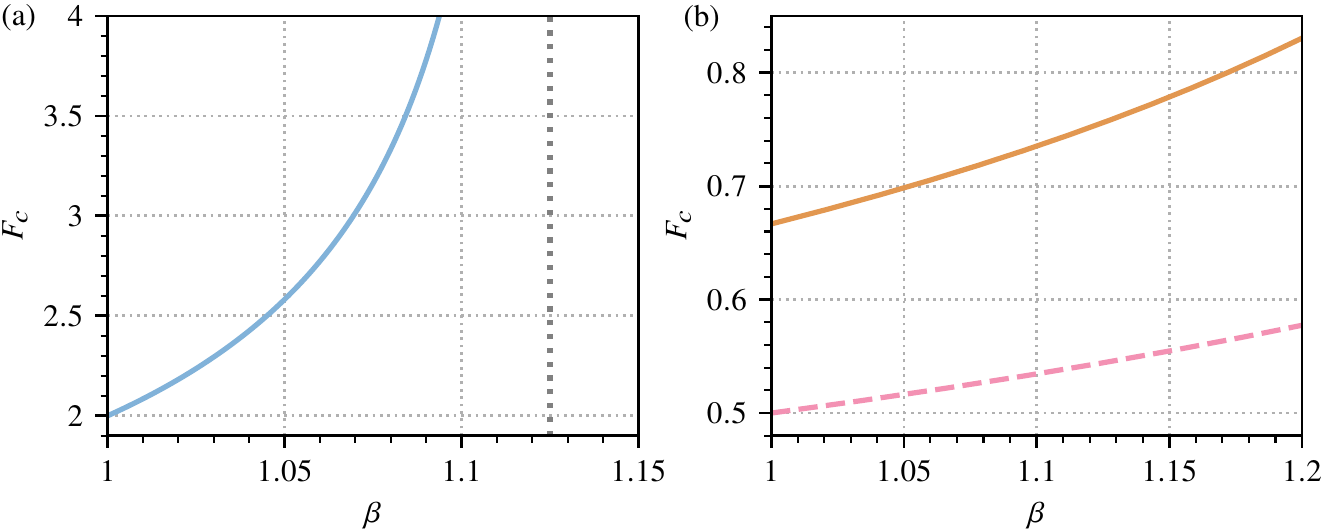}%
 \caption{%
     Dependence of the critical $\Fr = \Fr_c$ for instability of traveling
     waves on a constant ($h$- and $u$-independent) momentum shape factor $\beta$ for
     (a)~Ch\'ezy drag (b)~granular (solid) and viscous (dashed) drag.
     In panel~(a) the asymptote at $\beta = 1.125$ is also plotted (dotted).
 } 
 \label{fig:Frcrit beta}%
\end{figure}

To illustrate the effect that the momentum shape factor can have on the critical
$\Fr$, we consider the case where $\beta$ is approximated by a constant value.
Moreover, we suppose that the value of $\Fr_c$ is dictated by the upstream
regime and $\omega = -1$. (Both of these conditions are met in the cases of Ch\'ezy,
granular and viscous drag closures.)
Then, we Taylor expand Eq.~\eqref{eq:Frcrit} in powers of $(\beta - 1)$ to
obtain
\begin{equation}
    \Fr_c^2 = \Fr_{\mathrm{Tr}}^2 + (\beta - 1) \Fr_{\mathrm{Tr}}^3(2 +
\Fr_{\mathrm{Tr}}) + \ldots.
\label{eq:Frcrit Taylor ex}%
\end{equation}
Therefore, increasing $\beta$ raises the threshold for instability.
Indeed, the denominator in Eq.~\eqref{eq:Frcrit} vanishes in this case, if
\begin{equation}
    \beta = 1 + \frac{1}{\Fr_{\mathrm{Tr}} (\Fr_{\mathrm{Tr}} + 2)},
\label{eq:beta crit}%
\end{equation}
with $\Fr_c$ becoming unbounded as $\beta$ approaches this value.
In Fig.~\ref{fig:Frcrit beta}, we plot $\Fr_c(\beta)$ for our example drag
closures.
For Ch\'ezy drag [panel~(a)], $\Fr_c$ increases rapidly with $\beta$ and
diverges
at $\beta = 1.125$. The corresponding curves for granular and viscous
drag [panel~(b)] increase more steadily (though nevertheless significantly).
The qualitative difference between the effect of $\beta$ across the two panels is
encapsulated by Eq.~\eqref{eq:Frcrit Taylor ex}.

The stabilizing influence of larger $\beta > 1$ here may be explained within
the inequality given in~\eqref{eq:lam a}. It may be deduced in this case
that instability occurs when the bulk disturbance wave speed in the upstream
regime, $a_-$, surpasses the characteristic velocity $\lambda_{2-}$. Referring
to Eqs.~\eqref{eq:characteristics} and~(\ref{eq:eta a specific
all}\hyperref[eq:eta a specific all]{b}), we note that increasing the vertical
shear of the flow raises the speed of the characteristic while leaving $a_-$
unchanged.  This allows waves to remain stable at higher $\Fr$.  More
generally, the effect of $\beta(h,u)$ on the characteristics explains the
modified stability threshold of Eq.~\eqref{eq:Frcrit}.

\subsection{Implications for different solution regimes}
\label{sec:stability regimes}%
With the stability threshold determined, we may investigate which kinds of
solution are stable or unstable.  Firstly, we examine the stability of
continuous monoclinal waves versus shock solutions.

Inequality~\eqref{eq:lam a} allows us to prove an observation from
Sec.~\ref{sec:existence}, that the regime boundaries demarcating the onset of
shock development and existence of traveling wave solutions (see
Fig.~\ref{fig:existence}) must intersect at $(h_\infty, \Fr) = (1, \Fr_c)$.
Recall that these boundaries are given by the lines $\lambda_{2+} = 0$ and
$\lambda_{2-} = 0$ respectively.  When $h_\infty = 1$, we compute $a_- = a_+ = 1
- c_0(1) - \lim_{|\xi|\to\infty}(f_h / f_u)$, where $c_0$ is written as a
function of $h_\infty$, as in Sec.~\ref{sec:existence}.
We deduce via Eq.~\eqref{eq:c0}, that 
$c_0(1) = 1 + (\mathrm{d}u_\infty/\mathrm{d}h_\infty)|_{h_\infty=1}$.  
Therefore, by Eq.~\eqref{eq:fhfu lim}, we obtain
$a_\pm = 0$. This implies, by inequality~\eqref{eq:lam a},
that instability of the uniform layer occurs exactly when $\lambda_{2-} =
\lambda_{2+} = 0$, i.e.\ the intersection point of the two traveling wave regime
boundaries.

If the cases depicted in Fig.~\ref{fig:existence} (Ch\'ezy and granular drag)
are no more unstable in the downstream direction than upstream, it follows that
continuous monoclinal waves are always stable, because the curve bounding the
onset of shock development ($\lambda_{2+}=0$) is monotonic and bounded above by
the stability threshold of upstream flow, $\Fr = \Fr_c(1)$.
In the case of granular drag, we already observed in Fig.~\ref{fig:spec2},
that modes which are undamped upstream turn unstable at lower $\Fr_c$ than modes
which are undamped downstream.
More generally, 
if $\tau$ does not depend on any physical scales other than the flow height and
velocity, the local Froude number at which the downstream regime becomes
unstable must equal the critical Froude number of the upstream flow, $\Fr_c(1)$.
Since
the local Froude number in the downstream region differs from that of the
upstream region by a factor of $u_\infty / \sqrt{h_\infty}$,
we may write
\begin{equation}
    \Fr_c(h_\infty) = \min \left\{ \Fr_c(1), \Fr_c(1) \sqrt{h_\infty} / u_\infty
\right\}.
    \label{eq:Fc up down}%
\end{equation}
In this case, instability first occurs in the upstream flow if and only if
$\sqrt{h_\infty} / u_\infty > 1$.

If additional physical scales are present in the drag formulation, then
Eq.~\eqref{eq:Fc up down} cannot be used and the argument of Eq.~\eqref{eq:FTr
2} must be evaluated in both far-field regimes to determine which region is
more vulnerable to instability.  An example of a drag law that requires this
treatment is given by the resistance of a turbulent fluid in an open
rectangular channel of width $w$, which we non-dimensionalise with respect to
$H$, as in Eq.~(\ref{eq:variable subs all}\hyperref[eq:variable subs all]{c}).
Assuming the Ch\'ezy formula for turbulent bottom drag, then $\tau = (w +
2h)u^2 / (w + 2)$ (see e.g.  Refs.~\cite{Chow1959,Henderson1966} for details on
how to calculate such formulae).  Then, in either far field, using
$\tau(\vect{q}_0) = h_0$, we compute $(\sqrt{h_0} \mathrm{d}u_0 /
\mathrm{d}h_0)^{-1} = 2(w + 2h_0) \sqrt{h_0} / (u_0 w)$.  In the downstream
direction, $h_\infty / u_\infty^2 = (w + 2h_\infty)/ (w+2)$, which is strictly
less than one for $h_\infty < 1$.  Therefore, the downstream limit dictates the
evaluation of Eq.~\eqref{eq:FTr 2} and we determine the stability threshold 
\begin{equation}
    \Fr_{\mathrm{Tr}} = 2\left(1 + \frac{2h_\infty}{w}\right)
    \sqrt{\frac{w + 2h_\infty}{w + 2}}.
    \label{eq:FTr rect channel}%
\end{equation}
Note that the correction in Eq.~\eqref{eq:Frcrit} may be applied, if necessary,
to obtain the critical $\Fr$ for a general $\beta$.  
It is straightforwardly verified 
that this stability threshold could not have been obtained
from Eq.~\eqref{eq:Fc up down}. 
If $h_\infty = 1$,
Eq.~\eqref{eq:FTr rect channel} agrees with the critical Froude number
$\Fr_{\mathrm{Tr}}=2(1+2/w)$ for uniform layers in a rectangular
channel~\cite{Berlamont1981}.  When $h_\infty < 1$, the effective channel
breadth, measured with respect to the local flow height (i.e.\ $w/h_0$), is greater downstream
than upstream.
Therefore, since uniform layers in narrower channels are more stable, the
downstream flow turns unstable at lower $\Fr$, except in the limit $w\to\infty$,
where $\Fr_{\mathrm{Tr}}\to 2$ and both regions turn unstable at the same $\Fr$.

In the various cases where the drag law permits us to find a power law for the
velocity of a steady layer, $u_0 = h_0^m$, Eq.~\eqref{eq:Fc up down} does apply
and it is straightforward to see that the upstream flow becomes unstable at
lower $F$ than the downstream flow if and only if $m > 1/2$.
Here, Eq.~\eqref{eq:FTr 2} implies the following strikingly simple formula for
the linear stability threshold when $\beta(h,u)=1$,
%
%
%
%
\begin{equation}
    \Fr_c = \Fr_{\mathrm{Tr}} = \min_{h_0 = 1, h_\infty} \frac{1}{|m| h_0^{m-1/2}}.
    \label{eq:1/m result}%
\end{equation}
This result may be used to derive various well-known stability results, 
by selecting different $m$,
e.g.\ for Ch\'ezy, granular and viscous uniform layers, $u_0 = h_0^{1/2},
h_0^{3/2}, h_0^2$ and $\Fr_{\mathrm{Tr}} = 2$, $2/3$ and $1/2$
respectively~\cite{Jeffreys1925,Forterre2003,Trowbridge1987}.

We can now address whether continuous monoclinal waves (with $\beta = 1$) are
always stable in general for the class of drag laws with $u_0 = h_0^m$.
The upper bound of the continuous solution regime
($\lambda_{2+}=0$), may be rearranged to give the maximum $\Fr$ for continuous
waves, $\Fr = \Fr_{\mathrm{cont}}(h_\infty) = \sqrt{h_\infty}(1 - h_\infty) /
(1 - h_\infty^m)$.  This curve is strictly increasing  for $m \leq 3/2$.  It is
straightforward to show in this case, using Eq.~\eqref{eq:1/m result}, that
$\Fr_{\mathrm{cont}}(h_\infty) < \Fr_c(h_\infty)$, regardless of whether $m <
1/2$.  If instead, $m > 3/2$, then $\Fr_{\mathrm{cont}}(h_\infty)$ possesses a
turning point, which lies within the interval $0 < h_\infty < 1$, for all $m >
2$. It is in this latter case only, that continuous monoclinal waves can
suffer linear instability without first developing a shock. An example of a
system where continuous waves are not always stable is provided by the family
of `power law' fluids with drag formula $\tau(h,u) = (u/h)^n$, where $n>0$ is a
constant~\cite{Ng1994}. Then, $u_\infty(h_\infty) = h_\infty^{1+1/n}$, implying
that there exist continuous power-law waves with $n < 1$ (within a suitable
range of $h_\infty$) that become unstable prior to shock development.

Using the same assumptions as above, we investigate the stability of upturned
shock solutions (states with $h_0(0^-) > 1$).  The critical case, where shocks
are neither downturned, nor upturned, is given by $h_0(0^-) = 0$. Consulting
Eq.~\eqref{eq:h0minus}, we rearrange to obtain the curve
\begin{equation}
    \Fr = \Fr_{\mathrm{up}}(h_\infty) = 
    \frac{(1+h_\infty)^{\frac{1}{2}}(1-h_\infty)}{(2h_\infty)^{\frac{1}{2}}(1 -
h_\infty^m)}.
\label{eq:Frup}%
\end{equation}
Above this value of $\Fr$, solutions are upturned. 
We note, by L'H\^{o}ptial's rule, that 
$\Fr_{\mathrm{up}}$ also passes through $\Fr_c(1) = 1 / m$ at $h_\infty = 1$.
Moreover, assuming $m > 0$, we note that
$\Fr_{\mathrm{up}}(h_\infty) \to \infty$ as $h_\infty \to 0^+$.
The curve has a turning point within $0 < h_\infty < 1$ only if $m < 1/2$.
Therefore, all upturned shocks with $m \geq 1/2$ are unstable.
For $m < 1/2$, we must assess whether the existence of the turning point allows
$\Fr_{\mathrm{up}}(h_\infty)$ to drop below 
$\Fr_c(h_\infty)$ for any $0 < h_\infty < 1$.
In Appendix~\ref{appendix:upturned}, we show that it does not and hence we
conclude that upturned shock solutions are unstable for any $m$.

\section{Eigenmode structure}
\label{sec:eigenmode structure}%
In addition to the linear growth rate, $\sigma$, the spatial structure of the
corresponding modes plays a role in determining the evolution of disturbances.
We saw in Figs.~\ref{fig:spec1} and~\ref{fig:spec2} that the modes for granular
traveling waves are essentially oscillatory and can become
dramatically amplified at the wave front in some cases. 
In this section, we demonstrate that this amplification occurs for generic
choices of $\tau$ and show how it may be computed asymptotically when the
wavenumber is large.

Since the dominant disturbances are purely harmonic in at least one of the
far-field regions, we shall focus on modes which are undamped in the upstream
direction, i.e.\  $k_- \in \mathbb{R}$. It is straightforward to adapt
our analysis below to situations where the modes are undamped in the downstream.
The regimes of high and low wavenumber present tractable opportunities to
understand the spatial variation of modes. In the latter case, we note (as
discussed in Sec.~\ref{sec:linear stability}) that the higher of the two
branches of $\sigma$ passes through $\sigma = 0$ when $k_- = 0$.  By inspecting
Eq.~\eqref{eq:linear problem}, it is immediately clear that an $O(\epsilon)$
perturbation to $\sigma$ away from zero, does not alter $\hat{\vect{q}}_1$ to
leading order. In other words, at low $k_-$, modes are asymptotically close to
the neutral modes at $\sigma = 0$. This is evident, even in modes that are some
distance from the origin of the spectral plane. For example, the darker curve in
Fig.~\ref{fig:spec1}(e) clearly inherits its amplified peak at $\xi = 0$ from
the corresponding neutral mode $\hat{\vect{q}}_1 = (-h_0, -u_0)$, plotted in
Fig.~\ref{fig:spec1}(b).

Therefore, we proceed to the high-wavenumber limit, $k_- \gg 1$, which we have
shown above determines the most rapidly growing perturbations when the traveling wave
is unstable. This is a richer problem, which we find necessary to divide
into two cases, depending on the downstream depth.

\subsection{Finite downstream depth ($0 < h_\infty < 1$)}
\label{sec:short wavelength}%
First, suppose that the downstream height $h_\infty$ is finite and nonzero.
We analyze the spatial structure of $\hat{\vect{q}}_1$ by employing a WKB
approximation for the solution to Eq.~\eqref{eq:linear problem} (see e.g.\
Ref.~\cite{Bender1978} for details on this method). Setting $\varepsilon =
1/k_-$, we pose the ansatz
\begin{subequations}
\begin{gather}
    \hat{\vect{q}}_1(\xi) = \mathrm{e}^{\im \phi(\xi) / \varepsilon}
    \left[ \hat{\vect{r}}_0(\xi) + \varepsilon \hat{\vect{r}}_1(\xi) + \ldots
    \right],\label{eq:q asym}\\
    \sigma = \varepsilon^{-1}\sigma_0 + \sigma_1 + \ldots\label{eq:sigma asym}%
\end{gather}
\label{eq:wkb monoclinal}%
\end{subequations}
in the regime $\varepsilon \ll 1$, where the unknown functions $\phi$,
$\hat{\vect{r}}_0$, $\hat{\vect{r}}_1$ and the constants $\sigma_0$, $\sigma_1$ are
understood to be $O(1)$ with respect to $\varepsilon$.

The leading component of the growth rate in Eq.~\eqref{eq:sigma asym} is purely
imaginary and given by the system characteristics
[Eq.~\eqref{eq:characteristics}].  This is a fact straightforwardly verified by
substituting our expansions into Eq.~\eqref{eq:linear problem} and keeping only the
leading $O(\varepsilon^{-1})$ terms, to obtain
\begin{equation}
    (\phi' J - \im \sigma_0 I) \hat{\vect{r}}_0 = \vect{0}.
    \label{eq:O(1/eps)}%
\end{equation}
Therefore, $\im \sigma_0 / \phi'$ is an eigenvalue of the Jacobian, i.e.\ a
characteristic.
To match a purely harmonic disturbance upstream, we must impose $\phi \to \xi$
as $\xi \to -\infty$.  Therefore, $\im \sigma_0 =
\lim_{\xi\to-\infty}\lambda_{j}$, for either $j = 1$ or $2$, and we deduce that
\begin{equation}
    \phi'(\xi) = \frac{\lambda_j(1,1)}{\lambda_j(h_0(\xi), u_0(\xi))}.
    \label{eq:phi prime}
\end{equation}
This equation may be integrated (in principle) to obtain the frequency modulation 
of $\hat{\vect{q}}_1$, with respect to $\xi$.  The only
restriction on this procedure comes if the denominator in Eq.~\eqref{eq:phi
prime} vanishes.
This occurs if one of the characteristics changes sign. As argued in
Sec.~\ref{sec:existence}, only $\lambda_2$ can change sign along the wave and if
it does so, this happens across a shock at $\xi = 0$. 
Therefore, provided the hydraulic jump has finite depth, $\lambda_2$ is nonzero
either side of the shock.

The vector $\hat{\vect{r}}_0(\xi)$ is given by the eigenspace for the
characteristic curves as $J$ varies along the slope. It may be neatly expressed
using the characteristics themselves, as
\begin{equation}
    \hat{\vect{r}}_0(\xi) 
    = R(\xi)\begin{pmatrix}
        2h_0 \\
        B_1 \pm (\lambda_2 - \lambda_1)
    \end{pmatrix},
\end{equation}
where $R(\xi)$ is an unknown amplitude that we aim to determine.

\begin{figure}
 \centering%
    \includegraphics{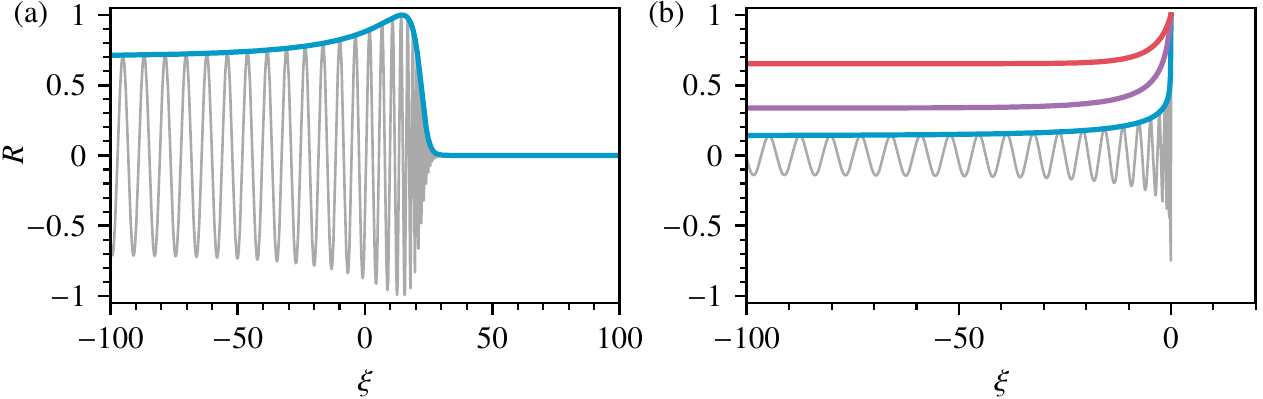}%
 \caption{%
 Spatial variation of the mode amplitude function $R$, determined numerically via
 Eq.~\eqref{eq:O(1) proj}, for granular traveling waves with $h_\infty
 = 0.5$. (a)~Amplitude $R(\xi)$ at~$\Fr = 0.5$ (blue), plotted alongside
 $u_1(\xi)$ (gray) for a spatially undamped mode with
 $\sigma = -0.014\,884 - \im$. (This growth rate corresponds to $k_- \approx
 0.74$. Higher wavenumber modes also agree well with $R$.)
 (b)~Mode amplitudes for waves in the discontinuous traveling wave regime,
 for $\Fr = 0.85$ (red), $0.7$ (purple) and $0.55$ (blue), approaching the 
 critical $\Fr \approx 0.547$, that marks the change between continuous and
 discontinuous states when $h_\infty = 0.5$ [see~Fig.~\ref{fig:existence}(b)].
 The $u_1$-field for a corresponding mode in the $\Fr = 0.55$ case, with $\sigma
 = -0.0113 -\im$ ($k_- \approx 0.85$) is also plotted (gray).
 } 
 \label{fig:R vs modes}
\end{figure}

At $O(1)$, we find
\begin{equation}
    (\im \phi' J + \sigma_0 I)\hat{\vect{r}}_1 = -J\hat{\vect{r}}_0' - (\sigma_1 I +
    N)\hat{\vect{r}}_0.
    \label{eq:O(1)}%
\end{equation}
To eliminate the unknown vector
$\hat{\vect{r}}_1$, we appeal to the eigenproblem adjoint to Eq.~\eqref{eq:O(1/eps)},
\begin{equation}
    \hat{\vect{l}}_0^T(i\phi'J + \sigma_0 I)^T  = \vect{0},
\end{equation}
%
and compute
\begin{equation}
    \hat{\vect{l}}_0(\xi) 
    = L(\xi) \begin{pmatrix}
        2(\Fr^{-2} + B_2) \\
        B_1 \pm (\lambda_2 - \lambda_1)
    \end{pmatrix}.
\end{equation}
We are free to choose the amplitude 
function $L(\xi)$
and do so according to the constraint
$\hat{\vect{l}}_0 \cdot \hat{\vect{r}}_0 = 1$. Then we project
Eq.~\eqref{eq:O(1)} onto $\hat{\vect{l}}_0$. By rearranging and setting
$\bar{\vect{l}}_0 = R \hat{\vect{l}}_0$ and $\bar{\vect{r}}_0 =
R^{-1}\hat{\vect{r}}_0$, we determine
\begin{equation}
    \lambda_j \frac{R'}{R} = -\sigma_1 - \bar{\vect{l}}_0 \cdot \left(
    J \bar{\vect{r}}_0' + N\bar{\vect{r}}_0 
    \right),\quad \mathrm{for}~j=1,2,
    \label{eq:O(1) proj}
\end{equation}
This is a first order differential equation for the mode amplitude function $R$.
Note that, as $\xi\to-\infty$, both $R'\to 0$ and $\bar{\vect{r}}_0' \to
\vect{0}$, so $\sigma_1 = -\lim_{\xi\to-\infty}\bar{\vect{l}}_0\cdot N
\bar{\vect{r}}_0$.  Since this expression is purely real, $\sigma_1$ must agree
with the high wavenumber growth rate formulae already computed in
Eqs.~\eqref{eq:s1 s2}. (Directly computing the matrix products confirms this.)
In Fig.~\ref{fig:R vs modes}(a) we plot a numerical solution to
Eq.~\eqref{eq:O(1) proj} for granular drag with $\Fr = 0.5$, $h_\infty = 0.5$.
It accurately captures the amplitude envelope of undamped eigenmodes, even those
with modest far-field wavenumber ($k_- \approx 0.74$ in the plot).

When written in full, the final two terms of Eq.~\eqref{eq:O(1) proj} are
complicated analytical expressions, which depend on $\vect{q}_0(\xi)$.
Nevertheless, we can gain insight into the spatial variation of $R$ by noting
that away from the upstream far-field, the right-hand side must deviate from
zero as the wave depth decreases (assuming $h_\infty < 1$). Therefore, if
$\lambda_j$ becomes small (compared with $\sigma_1$), we should expect to see
$R(\xi)$ grow rapidly (and exponentially) from its far-field value in the region of
the wave front.  Moreover, Eq.~\eqref{eq:phi prime} implies a corresponding
rapid frequency change.  There are two situations in which this can occur.
Firstly, in the case of discontinuous waves, $\lambda_2$ changes sign at the
shock. Therefore, if the shock depth is shallow, $\lambda_2$ is necessarily
small either side of $\xi=0$ and becomes zero in the limit $h_0(0^-) \to
h_\infty$.  We saw the effect of this in Fig.~\ref{fig:spec2}(b), which shows an
eigenmode for a wave that is very close to the regime boundary between
continuous and discontinuous solutions.  In Fig.~\ref{fig:R vs modes}(b), plot
$R(\xi)$ for $h_\infty = 0.5$ and $\Fr = 0.85$, $0.7$, $0.55$. As $\Fr$
approaches the regime boundary at $\Fr \approx 0.547$, $R(\xi)$ exhibits
progressively greater amplification at the shock.  
The second situation where $\lambda_2$ can become small, leading to similar
amplification, is in the limit $h_0,B_1\to 0$. That is, as the traveling wave
approaches a `flood wave' with dry downstream region. We consider the limiting
case in the proceeding section.

\subsection{Flood waves ($h_\infty=0$)}
\label{sec:flood waves}%
If $\beta(h,u) = 1$, flood waves suffer from a loss of strict hyperbolicity at
the front, where (recalling that $u_0 = c_0 = 1$ in this case) their
characteristics are identically zero. As argued in the previous subsection, this
renders inappropriate the standard WKB ansatz in Eq.~\eqref{eq:q asym}. Here, we
provide a separate analysis for this typical special case, in which the
underlying traveling wave equation~\eqref{eq:h0 ode} simplifies considerably to
\begin{equation}
    h_0' = \Fr^2 \left[ 1 - \frac{\tau(h_0)}{h_0}\right].
    \label{eq:h0 ode flood}%
\end{equation}
In the case of Ch\'ezy drag, a conceptually similar (though inequivalent)
equation was solved analytically by Bresse, who sought stationary steady
solutions to Eqs.~\eqref{eq:mass dim} and~\eqref{eq:mom dim} (i.e.\ $c_0 = 0$ and
$u_0 = 1/h_0$)~\cite{Bresse1859}. Bresse's solution in terms of elementary
functions is commonly available in hydraulics textbooks (see e.g.\
Refs.~\cite{Chow1959,Henderson1966}).  For traveling flood waves ($c_0 = 1$),
the profile differs. Specifically, we may integrate Eq.~\eqref{eq:h0 ode flood}
with $\tau(h_0) = u_0^2 = 1$, to obtain $h_0(\xi) = W[-\exp(\Fr^2 \xi - 1)] +
1$, where $W$ is Lambert's function. Equation~\eqref{eq:h0 ode flood} has also
been studied in the context of granular
avalanches~\cite{Pouliquen1999a,Gray2009} and likewise admits an analytical
solution~\cite{Gray2009}.

A dry downstream region places some extra constraints on the linear stability
formulation posed in Sec.~\ref{sec:linear stability}. The traveling wave
velocity must equal the flow velocity at the front, so $c_1 = u_1(0^-)$.
Furthermore, there can be no disturbance for $\xi > 0$, so $h_1(0^+) = u_1(0^+)
= 0$. Finally, in order for the perturbation to be considered small in the
front region, $h_1(0^-) = 0$ and we must check that $h_1 \lesssim h_0$ as $h_0
\to 0$.

Depending on the drag law, $h_0$ varies differently in the front region.
We encompass these different profiles by writing the general expansion
%
\begin{equation}
\tau(h_0,u_0)=\Lambda(h_0, u_0) h_0^\delta+O(h_0^{\delta+1}), 
\label{eq:tau0 expansion}%
\end{equation}
where
$\Lambda(h_0,u_0)$ is finite and non-vanishing as $h_0\to 0$ and $\delta$ is an
arbitrary exponent. For our three main example closures, $\delta = -1$
(viscous), $\delta = 0$ (Ch\'ezy) and $\delta = 1$ (granular).  Substituting
Eq.~\eqref{eq:tau0 expansion} into Eq.~\eqref{eq:h0 ode flood} and simplifying yields 
\begin{equation}
h_0' = 
\Fr^2(1 - \Lambda h_0^{\delta - 1}) + \ldots.
\label{eq:h0 ode expan}%
\end{equation}
For any $\delta > 1$ the last term is subdominant at the front and
the resulting equation leads to negative flow depths. Therefore, we discount
these cases.  Otherwise, Eq.~\eqref{eq:h0 ode expan} may be integrated to give
\begin{equation}
h_0=\left(-A\xi\right)^{\gamma}+\ldots\qquad\hbox{when}\qquad -1 \ll \xi < 0,
\label{eq:h0 expansion}%
\end{equation}
where $\gamma=1/(2-\delta)$ and 
\begin{equation}
    A= \begin{cases}
        (2-\delta)\Fr^2[\Lambda(0,1) - 1] & \mathrm{if~}\delta = 1,\\
        (2-\delta)\Fr^2\Lambda(0,1) & \mathrm{if~}\delta < 1.
        \end{cases}
\end{equation}
On substituting Eqs.~\eqref{eq:tau0 expansion} and~\eqref{eq:h0 expansion} into
the the linear stability problem, Eq.~\eqref{eq:linear problem}, it may be
deduced that $h_1$ is at most order $(-\xi)^\gamma$ at the front.
The details are given in Appendix~\ref{app:h1 check}.
Therefore,  $\varepsilon |h_1| \ll h_0$, for sufficiently small $\varepsilon > 0$,
implying that solutions to Eq.~\eqref{eq:linear problem} are indeed linear
perturbations in the case of flood waves.

To capture the structure of these modes at high wavenumber, we employ a slightly
modified version of Eqs.~(\ref{eq:wkb monoclinal}\hyperref[eq:wkb
monoclinal]{a,b}). Since only one far-field direction is relevant
for the linear flood wave problem, we shall write $k\equiv k_-$.  At leading
order in $\varepsilon = 1/k \ll 1$, we suppose
\begin{subequations}
\begin{gather}
    \hat{\vect{q}}_1(\xi)
    =
    R(\xi){\rm e}^{\im\phi(\xi)/\varepsilon} 
    (\hat{\vect{r}}_0 + \varepsilon \hat{\vect{r}}_1 + \ldots)
+
S(\xi){\rm e}^{-\im\phi(\xi)/\varepsilon}
(\hat{\vect{s}}_0 + \varepsilon \hat{\vect{s}}_1 + \ldots),\label{eq:wkb flood 1}\\
\sigma = \varepsilon^{-1}\sigma_0 + \sigma_1 + \ldots,
\label{eq:wkb flood 2}%
\end{gather}
\end{subequations}
where $R(\xi)$ is finite and non-vanishing and $S(\xi)$ vanishes as
$\xi\to-\infty$.  The conjugate term in Eq.~\eqref{eq:wkb flood 1} is necessary in
this case to correctly represent the mode towards the front region -- a point
which will be clarified later. 
Linear independence of these terms means that the
analysis to determine $\hat{\vect{r}}_0, \hat{\vect{r}}_1, \ldots$ is identical
to the presentation in Sec.~\ref{sec:short wavelength}.
However, as anticipated, evaluating Eq.~\eqref{eq:phi prime} and integrating gives
\begin{equation}
    \phi(\xi)=\int^{\xi}_0\frac{1}{\sqrt{h_0(s)}}\;{\rm d}s,
    \label{eq:phi}%
\end{equation}
meaning that $\phi$ can no longer be considered $O(1)$ with respect to
$\varepsilon$.  Nevertheless, this integral is guaranteed to converge, since
$\gamma \leq 1$.  Therefore, we are only required to modify our asymptotic
expansion for $\hat{\vect{q}}_1(\xi)$ close to the front.  Proceeding with the
WKB analysis, we determine via Eq.~\eqref{eq:O(1) proj}, that 
\begin{equation}
\frac{R'}{R}=-\frac{
2\Fr^2 f_h h_0+3 h_0'
\pm2\Fr(2\sigma_1+f_u)\sqrt{h_0}
}{4h_0}.
\label{Rflood}
\end{equation}
Since the mode amplitude is constant in the far field, we write $R\to R_-$ as
$\xi\to-\infty$, where $R_-$ is to be determined. On integrating
Eq.~\eqref{Rflood}, we find
\begin{align}
R(\xi)h_0(\xi)^{3/4}&=R_-\exp
\left(-\frac{\Fr^2}{2}\int_{-\infty}^\xi
    f_h 
    \pm 
\frac{%
    2\sigma_1+f_u
}{\Fr\sqrt{h_0}}{\rm d}\acute\xi \right),
\label{eq:outer}%
\end{align}
where $\sigma_1$, $f_u$, $f_h$ and $h_0$ are understood to be integrated with
respect to the dummy slope variable $\acute\xi$.
The integrand in Eq.~\eqref{eq:outer} vanishes in the far field in order
to satisfy the boundary condition there. (This may be separately confirmed by
computing $\sigma_1$.)
Therefore, we appeal to the near-front expansions in Eqs.~\eqref{eq:tau0 expansion}
and~\eqref{eq:h0 expansion} and deduce that the dominant part of the integral is 
\begin{equation}
    \exp\left(-\frac{\Fr^2}{2}\int_{-\infty}^\xi f_h \mathrm{d}\acute\xi\right) 
    =
    \exp\left[
    \frac{1}{2}
    \int^1_{h_0}
    \frac{h_0^{\delta-1}}{h_0 - \Lambda(h_0,1)h_0^{\delta}}
    \left(
    \frac{\partial \Lambda}{\partial h_0}h_0
    + (\delta - 1) \Lambda(h_0,1)
    \right)
    \mathrm{d}h_0
\right]
    \sim 
    h_0^{\frac{\delta}{2} - \frac{1}{2}}
    \label{eq:outer scaling}%
\end{equation}
to leading order.  If $\delta = 1$, we must require that $\partial \Lambda /
\partial h_0$ is nonzero. However, we can see from Eq.~\eqref{eq:h0 ode expan}
that this is necessary for a front to form in the first place.  Therefore, from
Eq.~\eqref{eq:outer}, we deduce that $R(\xi) \sim h_0^{(2\delta - 5)/4}=
h_0^{-(2+\gamma)/(4\gamma)}$ at the front.  Since $0 < \gamma \leq 1$, it
diverges at least as rapidly as $h_0^{-3/4}$ there. Consequently, the WKB
approximation cannot attain the boundary condition $(h_0, u_0) = (0, 1)$ at the
front and as expected, we must consider this region separately.

To examine this inner region it is convenient to write the eigenmode
equations~\eqref{eq:linear problem} as a single second order equation for $u_1$.
We obtain 
\begin{equation}
h_0 u_1'' + (2h_0' + \Fr^2 f_h h_0)u_1' - \Fr^2 \sigma(\sigma + f_u) u_1 = 0.
\label{eq:u1 2nd order}%
\end{equation}
%
By evaluating~\eqref{eq:phi} near the front we note that $\phi \sim \varepsilon$
when $\xi \sim \varepsilon^{2/(2-\gamma)}$. Consequently we set
$\eta=-\xi/\varepsilon^{2/(2-\gamma)}$ in order to capture this inner scale. We
may write the spatial derivatives, $h_0$, $h_0'$, $f_h$ and $f_u$ in terms of
$\eta$, once again making use of our expansions in Eqs.~\eqref{eq:tau0
expansion} and~\eqref{eq:h0 expansion}.  On substituting these into
Eq.~\eqref{eq:u1 2nd order} and making use of our expansion for $\sigma$ in
Eq.~\eqref{eq:wkb flood 2} [note that $\sigma_0^2 =
-1/\Fr^{2}$ may be determined from Eqs.~\eqref{eq:O(1/eps)} and~\eqref{eq:phi prime}], the leading part of the resulting equation is found to
be $O(\varepsilon^{-2})$. It reads
\begin{equation}
\frac{\mathrm{d}^2 u_1}{\mathrm{d}\eta^2}
+\frac{1+\gamma}{\eta}
\frac{\mathrm{d} u_1}{\mathrm{d}\eta}
+\frac{1}{(A\eta)^\gamma}
u_1=0.
\end{equation}
%
%
%
The solution that passes through $u_1(0) = 1$ is given by
\begin{equation}
u_1(\eta)=\Gamma\left(\frac{2}{2-\gamma}\right)Y^{\gamma/(\gamma-2)}J_{\gamma/(2-\gamma)}\left(2Y\right),
\label{eq:inner}%
\end{equation}
where $Y=\eta(2-\gamma)^{-1}(A\eta)^{-\gamma/2}$, $\Gamma$ is the gamma function and
$J_n$ denotes the Bessel function of the first kind of order $n$.  In the far
field $\eta\gg 1$, we find
\begin{equation}
u_1(\eta)\sim \frac{1}{\pi^{1/2}}\Gamma\left(\frac{2}{2-\gamma}\right)Y^{(\gamma+2)/(2\gamma-4)}
\cos\left(2Y -\frac{(\gamma+2)\pi}{4(2-\gamma)}\right).
\label{eq:inner far}%
\end{equation}

We note that $Y^{(\gamma + 2)/(2\gamma - 4)} \sim (-\xi)^{-(2+\gamma)/4} \sim
h_0^{-(2+\gamma)/(4\gamma)}$, in agreement with the near-front scaling of the
outer WKB approximation.  Moreover, by substituting $h_0 = (-A\xi)^\gamma$ into
Eq.~\eqref{eq:phi} and integrating, it is straightforward to show that $2Y =
\phi / \varepsilon$.  Therefore, as $\eta$ becomes large, the decaying
oscillations of Eq.~\eqref{eq:inner}, match the leading-order behavior of the
outer solution constructed in Eq.~\eqref{eq:wkb flood 1}, as $\xi$ becomes
small.  It was for the purpose of matching the cosine function in
Eq.~\eqref{eq:inner far} that we included the conjugate term in the WKB ansatz.
However, since $S(\xi)\to 0$ as $\xi\to-\infty$, we concentrate on determining $R_- =
\lim_{\xi\to-\infty} R(\xi)$. By appealing
to Eqs.~\eqref{eq:outer}, ~\eqref{eq:outer scaling} and~\eqref{eq:inner far}, we
establish that
\begin{equation}
|R_-|
=
\frac{
\left[
    (2-\gamma)\varepsilon A
\right]^{(\gamma+2)/(4 - 2\gamma)}
}{2\pi^{1/2}}\Gamma\left(\frac{2}{2-\gamma}\right)
\exp(\mathcal{I}),
\label{eq:R minus}%
\end{equation}
where
\begin{equation}
    \mathcal{I} = \int_0^1
\frac{
    \Fr f_h h_0\pm
    (2\sigma_1+f_u)h_0^{1/2}
    }{2\Fr (\tau_0 - h_0)}
+
\frac{(1-\gamma)}{2\gamma h_0}{\rm d}h_0.
\label{eq:I int}%
\end{equation}

\begin{figure}
 \centering%
    \includegraphics{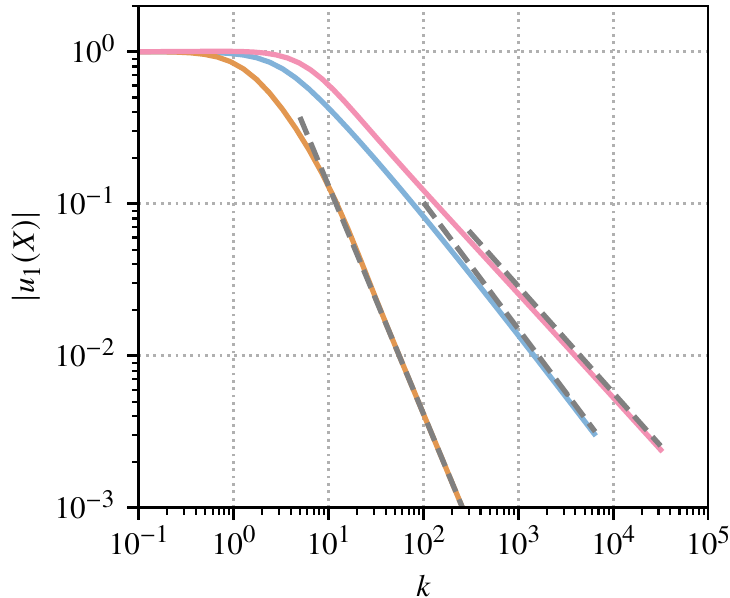}%
 \caption{%
     Far-field amplitudes of (the dominant branch of) marginal eigenmodes, for $\Fr = 2.5$ flood
     waves with (a)~Ch\'ezy (blue), (b)~granular (orange) and (c)~viscous (pink)
     drag.  For each drag law, we compute $u_1$ from Eq.~\eqref{eq:linear
     problem} and evaluate its amplitude at $\xi = X = -40$, which (at the given
     $\Fr$) is sufficiently far upslope that the underlying traveling wave
     approximates a uniform layer.  Also plotted in each case is a section of
     the corresponding asymptotic amplitude $|R_-|$ (dashed gray), given
     by~Eq.~\eqref{eq:R minus}, verifying convergence as $k\to\infty$. 
 } 
 \label{fig:amp scalings}
\end{figure}

Equations~\eqref{eq:R minus} and~\eqref{eq:I int} complete the matched expansion
for marginal flood wave stability modes at high wavenumber $k=1/\varepsilon$.
Note that the two expressions for the integral in Eq.~\eqref{eq:I int}
correspond to two independent branches of modes with asymptotic growth rate
$\sigma_1 = -\lim_{\xi\to-\infty} (f_u \pm F f_h) / 2$, as determined by the
formulae in Eqs.~\eqref{eq:s1 s2} (or the analysis of
Sec.~\ref{sec:short wavelength}).  The amplitude $|R_-|$ dictates the asymptotic
decay of $u_1(\xi)$ in the far field, or equivalently, the growth of the
perturbation at the front.  Noting that $1/|R_-| \sim k^{(\gamma + 2)/(4 -
2\gamma)}$ for $k\gg 1$, we see that disturbances ultimately become severely
amplified from tail to front as $k\to\infty$, irregardless of the drag
formulation, since
\begin{equation}
    k^{1/2} < k^{(\gamma + 2)/(4 - 2\gamma)} \leq k^{3/2}.
\end{equation}
In nature, this amplification is not truly unbounded, since we expect it to be
ultimately damped by physical processes omitted from the governing equations
that are only relevant at small length scales (e.g.\ turbulent diffusion).
Nevertheless, this divergent behavior dictates the spatial properties of modes
at finite, but large wavenumber.
For our example drag formulations, $1/|R_-| \sim k^{7/10}$ (viscous),
$k^{5/6}$ (Ch\'ezy) and $k^{3/2}$ (granular).
We verify the asymptotic formula of Eq.~\eqref{eq:R minus} numerically in
Fig.~\ref{fig:amp scalings},
by plotting the far-field amplitude of $u_1(\xi)$ for the dominant (higher
$\sigma_1$) branch of
modes, over a range of $k$. This
agrees well with $|R_-|$ as $k\to\infty$.
The predicted scaling emerges early on, particularly in the granular case,
where the modes have decayed to one tenth of their original amplitude by $k
\approx 5$.

\section{Discussion}
\label{sec:discussion}%
We have presented detailed analysis of the linear stability properties of
monoclinal traveling waves in shallow flows of arbitrary rheology.  By
considering a setting that encompasses a broad family of flow models, the
essential differences separating the properties of various flows are elucidated.  In
table~\ref{tab:summary} we summarize some of our conclusions for different drag
parametrisations, in the case where the momentum shape factor $\beta(h,u) = 1$. 
\begin{table}
    \begin{ruledtabular}
\begin{tabular}{ccccccc}
    ~ & ~ & \multicolumn{3}{c}{$0 < h_\infty < 1$} &
    \multicolumn{2}{c}{$h_\infty = 0$} \\
    \cline{3-5}
    \cline{6-7}
    Drag & $\Fr_{\mathrm{Tr}}$ & $u_\infty(h_\infty)$ & Shock before
    instability? & Upstream unstable first? & Front scaling & $1/|R_-|$ \\
    \colrule
    $u^2$ & $2$ & $h_\infty^{1/2}$ & Yes & No ($u_\infty / \sqrt{h_\infty} =
    1$)${}^\dagger$ & $(-\xi)^{1/2}$ & $\sim k^{5/6}$ \\
    $\mu(h^{3/2}/u)h$ & $2/3$ & $h_\infty^{3/2}$ & Yes & Yes & $(-\xi)$ & $\sim k^{3/2}$ \\
    $(u/h)^n$ & $n / (n + 1)$ & $h_\infty^{1 + 1/n}$ & Iff $n > 1$ & Yes ($n >
    0$) & $(-\xi)^{1/(2+n)}$ & $\sim k^{\frac{5+2n}{6+4n}}$ \\
    $u^a h^b$ & $|a| / |1-b|$ & $h_\infty^{(1-b)/a}$ & Iff $2a + b > 1$ & Iff $a/2
    + b > 1$ & $(-\xi)^{1/(2-b)}$ & $\sim k^{\frac{5-2b}{6-4b}}$ \\
\end{tabular}
    \caption{Summary of the stability properties for different
        drag laws, in the case $\beta(h,u) = 1$. (${}^\dagger$As indicated, in
        this case the upstream and downstream Froude numbers are identical.
        However, in Sec.~\ref{sec:stability regimes}, we showed that the
        downstream turns unstable first if this drag formulation is generalized
        to include the effect confining the flow within a finite rectangular
        channel.)
    }
    \label{tab:summary}%
    \end{ruledtabular}
\end{table}
In addition to the example drag laws used throughout the text, we include the
properties of the general function $\tau(h,u) = u^a h^b$, where $a$ and $b$ are
arbitrary constants. This parametrisation encompasses Ch\'ezy, viscous, power
law drag and many other possible drag laws. Its listed properties may be readily
determined using the results of Secs.~\ref{sec:linear stability}
and~\ref{sec:eigenmode structure}.

As shown in Sec.~\ref{sec:linear stability}, the linear stability threshold for
traveling waves is ultimately given by Trowbridge's
criterion~\eqref{eq:trowbridge} applied to the upstream and downstream regions,
plus a correction if $\beta$ deviates from unity, given by
Eqs.~\eqref{eq:Frcrit} and~\eqref{eq:Frcrit Taylor ex}.  The corresponding
critical $\Fr$ (labeled $\Fr_{\mathrm{Tr}}$) is provided in the second column
of table~\ref{tab:summary}.  The third column gives the dependence of the
downstream velocity on the flow height. It highlights the simpler linear
stability relationship, derived in Eq.~\eqref{eq:FTr 2}, where
$\Fr_{\mathrm{Tr}}$ is given by the inverse of the exponent in the formula
$u_\infty(h_\infty) = h_\infty^m$ and more generally, by the derivative of the
steady velocity of a uniform layer with respect to its depth, evaluated in the
appropriate far-field limit.  
For the general drag $\tau(h,u) = u^a h^b$, the exponents $a \neq 0$, $b = 1$
are interesting to consider, since $\Fr_{\mathrm{Tr}} \to\infty$ as $b\to 1$.  A
physical interpretation of this situation comes from the paper of Trowbridge,
who showed how to reformulate inequality~\eqref{eq:trowbridge} as an energy
stability criterion~\cite{Trowbridge1987}.  When specialized to the present
case, that analysis shows that the rate of total energy input to an
infinitesimal disturbance from gravitational forcing is always exceeded by the
corresponding rate of energy loss due to work done by bottom stresses and
consequently, linear perturbations can only decay (regardless of $\Fr$).  Such a
$\tau$ arises in the limit of the drag from a turbulent rectangular channel
considered in Sec.~\ref{sec:stability regimes}, as the channel width tends to
zero ($a = 2$, $b = 1$). This limit has a practical application in modeling the
flow of turbulent gravity currents through densely packed obstacles such as
vegetated areas~\cite{Hatcher2000,Nepf2012}, but the relevant shallow water
linear stability problem has not previously received attention.

Our analysis of linear stability thresholds may be adapted straightforwardly
for the case of purely downslope perturbations of two-dimensional traveling
waves which are uniform along the perpendicular cross-slope direction.  However,
recent work proves that Trowbridge's criterion is violated for some rheologies
when non-slope-aligned disturbances are accounted for~\cite{Zayko2019}.
Therefore, a fully two-dimensional extension of our results would require some
care.
Furthermore, as discussed in Sec.~\ref{sec:linear stability}, our analysis only
strictly covers the stability with respect to modes within the continuously
parametrised essential spectrum of the linear problem. Consequently, it may be
the case for some drag formulations, that unstable isolated modes exist at
lower $\Fr$ than predicted here.  This is important to consider. For example,
in thin-film flows over an inclined plane controlled by viscous and capillary
processes, discrete modes can drive instability
of the contact line~\cite{Ye1999}.  While numerical techniques exist that could
rule out this possibility for specific instances of our 
setting~\cite{Barker2018}, obtaining general analytical results presents a far greater
challenge.  Nevertheless, recent studies rule out destabilization by
discrete modes for Ch\'ezy solutions, ultimately leading to a proof that
traveling waves with $\Fr < 2$ are nonlinearly
stable~\cite{Yang2020,Sukhtayev2020}. Therefore, there is reason to hope that
these efforts could be extended for other drag laws, or even a general drag
term.

Also of interest are results concerning the properties of traveling wave
solutions. In Sec.~\ref{sec:existence}, we derived the regime of existence for
continuous waves and their degeneration into discontinuous states.  In
the cases of Ch\'ezy and granular drag, increasing $\Fr$ for continuous states
always leads to formation of a shock, prior to development of linear
instability.  Later, in Sec.~\ref{sec:stability regimes}, a condition for the
existence of unstable continuous solutions was obtained and is applied in
table~\ref{tab:summary} (column four).

Traveling waves turn linearly unstable as their eigenmodes cross into the
positive half-plane (see Fig.~\ref{fig:spec2}). In Sec.~\ref{sec:linear
stability}, various spectra for granular waves were computed numerically and it
was observed that the onset of instability was dictated by the destabilization
of modes that are only nonzero upstream ($\xi < 0$). The far-downstream region
remains stable until $\Fr$ reaches a higher value. The respective vulnerability
of the far-field regimes depends on their local Froude numbers and is thus
easily determined (see also Sec.~\ref{sec:stability regimes}). Analogous
criteria for other drag laws are summarized in column five of
table~\ref{tab:summary}. 

We plotted many example eigenmodes throughout the paper. They typically possess
the intriguing property that their spatial frequency varies between limiting
values up- and downstream.  The dominantly growing modes within unstable regimes
are spatially undamped in at least one of the far field directions and have
asymptotically high wavenumber.  Consequently, we have analyzed in detail this
special class of modes.  In Sec.~\ref{sec:eigenmode structure}, we derived
equations for the variation of their amplitude and frequency in space.  Both of
these quantities are often strongly amplified near the front of the wave. Most
pressingly, the mode amplitude diverges for waves at the boundary between
continuous and discontinuous states, and for waves with zero depth downstream.
The latter case seems particularly important, due to the physical significance
of flood wave states and the fact that divergence occurs independently of $\Fr$.
In Sec.~\ref{sec:flood waves}, we showed that the rate of this divergence as the
wavenumber increases, depends on the spatial variation of the underlying
traveling wave close to the front. Granular waves, whose depth varies linearly
with $\xi$ at the front, diverge the most severely. (Drag formulations that
would lead to stronger divergence are unable to form steady flood wave
solutions.) Results for other example drag laws are summarized are the final two
columns of table~\ref{tab:summary}, and we note that the complete asymptotic
dependence for an arbitrary drag law is developed in Sec.~\ref{sec:flood waves}.
An interesting open question is: to what extent does the extreme amplification
of linear modes affect the nonlinear development of instabilities? While it
seems likely for any reasonable drag formulation that roll waves emerge within
uniform depth regions, there is also the potential for the front to become
disrupted and even for the wave to split in two. Our analysis suggests that
granular waves might be particularly vulnerable to disruption.

Though our findings have typically been illustrated using examples with
$\beta(h,u) = 1$, we have largely been able to present results for the case of
a general momentum shape factor closure. Increasing $\beta$ to a constant value
above unity acts to stabilize traveling waves and can have a significant effect
on the linear stability threshold. This has been noted before in the cases of
Ch\'ezy~\cite{Berlamont1981} and power law drag~\cite{Ng1994}. For example,
raising $\beta$ to $1.05$ in a turbulent (Ch\'ezy) fluid increases $\Fr_c$ from
$2$ to roughly $2.6$.  This suggests that when high fidelity calculations or
simulations are called for, a suitable model for $\beta$ should be employed.
The effects that variation of $\beta$ with $h$ and $u$ have are nontrivial to
tease apart with generality. Nevertheless, we included the derivatives $\partial
\beta / \partial h$ and $\partial \beta / \partial u$ in our analysis, since
even simple formulae for $\beta$ might introduce this dependency, e.g.\ if
shear is modeled as a function of flow Reynolds number.  When including these
terms, we find that it is important to avoid situations where the system could
lose hyperbolicity (leading to an ill-posed initial value problem), which occurs
if Eq.~\eqref{eq:ill p} is satisfied.  Otherwise, we expect the effects of
$\beta$ to be essentially quantitative (though nevertheless important).  Varying
$\beta$ does not appear to greatly affect the character of the wave profiles,
except in cases wherein (fixing all other parameters) it causes solutions to
cross a regime boundary, e.g.\ by converting a continuous state into a shock
(see Fig.~\ref{fig:existence}).  Moreover, we note that shear in the flow
profile does not affect the fact of mode amplification near the front regions of
certain waves, despite its necessary influence on the details of the linear
problem.  The precise effects of momentum shape factor closures in individual
models and on the linear and nonlinear development of instabilities offer
interesting avenues for future research.

\acknowledgments{
We thank D.\ Barkley, L.\ T.\ Jenkins, C.\ G.\ Johnson, J.\ C.\ Phillips, L.\
S.\ Tuckerman and M.\ J.\
Woodhouse for valuable discussions. This research was supported by the
Newton fund grant NE/S00274X/1 and the Royal Society grant APX/R1/180148.
}

\appendix

\section{Upturned shocks are always unstable}
\label{appendix:upturned}%
In this Appendix, we show that $\Fr_{\mathrm{up}}(h_\infty) > \Fr_c(h_\infty) =
h_\infty^{1/2-m}/m$ for all $0 < h_\infty < 1$ and $m < 1/2$, given the
assumptions $\beta(h,u) = 1$, $u_\infty = h_\infty^m$ and $m > 0$.  This
completes the argument of Sec.~\ref{sec:stability regimes} that upturned shock
solutions are unstable in this case, irregardless of the drag law. By consulting
Eq.~\eqref{eq:Frup} and rearranging $\Fr_{\mathrm{up}}(h_\infty) >
\Fr_c(h_\infty)$, we note that it
is equivalent to confirm that
\begin{equation}
    G(h_\infty; m) = \frac{(1-h_\infty)(1+h_\infty)^{1/2}}{\sqrt{2}}
    -\frac{h_\infty^{1-m}-h_\infty}{m} > 0.
    \label{eq:G ineq}%
\end{equation}
If $m = 1/2$, then
\begin{equation}
    G(h_\infty; m) = (1 - h_\infty^{1/2}) \left[
        (1 + h_\infty^{1/2})(1 + h_\infty)^{1/2}
        - 2\sqrt{2} h_\infty^{1/2}
    \right]
\end{equation}
and it is straightforwardly shown, e.g.\ via analysis of the term within the
square brackets, that this function is strictly positive for $0 <
h_\infty < 1$.
Furthermore, we compute the derivative
\begin{equation}
    \frac{\partial G}{\partial m}
    =
    \frac{h_\infty}{m^2}\left[
        \frac{1}{h_\infty^m}
        \left(
        1 + m \log h_\infty
        \right)
        -1
    \right],
\end{equation}
which has zeros only at $h_\infty = 0, 1$ and is strictly negative for $0 <
h_\infty < 1$. Therefore, decreasing $m$ from $1/2$, increases $G(h_\infty; m)$
for any such $h_\infty$, implying that inequality~\eqref{eq:G ineq} holds for
any $m < 1/2$, as required.

\section{Flood wave near-front perturbations}
\label{app:h1 check}%
In the special case of flood waves ($h_\infty = 0$) and $\beta = 1$, studied in
Sec.~\ref{sec:flood waves}, it is necessary to check that $\hat{\vect{q}}_1$ can
legitimately be considered as a linear perturbation at the front, where $h_0 \to
0$.

Since $u_0 = c_0 = 1$ for flood waves, the matrices $J$ and $N$ [defined in
Eqs.~\eqref{eq:jacobian} and~\eqref{eq:N}] are greatly simplified. We
compute
\begin{equation}
    J = 
    \begin{pmatrix}
        0 & h_0 \\
        \Fr^{-2} & 0
    \end{pmatrix},
    \quad
    \mathrm{and}
    \quad
    N = 
    \begin{pmatrix}
        0 & h_0' \\
        f_h & f_u
    \end{pmatrix}.
\end{equation}
By making use of the near-front of expansions in Eqs.~\eqref{eq:tau0 expansion}
and~\eqref{eq:h0 expansion}, we may then write down the linear problem in the
front region
\begin{subequations}
\begin{align}
    -\Fr^{-2} h_1' &= 
    \left[ 
        (\delta - 1)\Lambda(\vect{q}_0) + 
        (-A\xi)^{\gamma} \frac{\partial \Lambda}{\partial
        h}\bigg|_{\vect{q}=\vect{q}_0}
    \right] (-A\xi)^{-1} h_1
    + 
    \left[
        \sigma + (-A\xi)^{\gamma - 1} \frac{\partial \Lambda}{\partial
        u}\bigg|_{\vect{q} = \vect{q}_0}
    \right] u_1 + \ldots, \label{eq:h1 flood wave ode}\\
    -u_1' &= 
    \sigma (-A\xi)^{-\gamma} h_1
    +
    \gamma A (-A\xi)^{-1} (c_1 - u_1) + \ldots.
\end{align}
\label{eq:h1u1 flood wave ode}%
\end{subequations}
Then we use the fact that $0 < \gamma \leq 1$ (using $\delta \leq 1$ and the
definition of $\gamma$ from Sec.~\ref{sec:flood waves}) to deduce that the
$\partial \Lambda / \partial h$ term of Eq.~\eqref{eq:h1 flood wave ode} is
subdominant and may be neglected.

A suitable expansion for the perturbations that satisfies the front conditions
$h_1(0^-) = 0$, $u_1(0^-) = c_1$, is 
\begin{equation}
h_1 = K_1 (-\xi)^{\alpha_1} + \ldots, \quad
u_1 = c_1 + K_2 (-\xi)^{\alpha_2} + \ldots,
\end{equation} 
where $\alpha_1$, $\alpha_2$, $K_1$ and $K_2$ are constants to be determined.
On substituting these into Eqs.~(\ref{eq:h1u1 flood wave ode}\hyperref[eq:h1u1
flood wave ode]{a,b}) we immediately see that, for the remaining terms to
balance at leading order, $\alpha_1 = \gamma$ and $\alpha_2 = 1$.  Therefore,
since $h_0 = (-A\xi)^\gamma + \ldots$ at the front, this means that $h_1
\lesssim h_0$ in the front region, as required.

\end{document}